\documentclass[10pt]{iopart}  
\usepackage{iopams}

\usepackage{amsfonts}
\usepackage{epsf}
\usepackage{euscript}
\usepackage{graphicx}
\usepackage{color}

\begin{document}

\title[Asymptotic Solutions of the Phase Space Schr\"{o}dinger Equation]{Asymptotic Solutions of the Phase Space Schr\"{o}dinger Equation: Anisotropic Gaussian Approximation}

\author{Panos D. Karageorge$^{1,2}$ and George N. Makrakis$^{1,2,3}$\footnote{\texttt{pkaragiorgos}@\texttt{tem.uoc.gr} and \texttt{makrakg}@\texttt{tem.uoc.gr}, \texttt{makrakg}@\texttt{iacm.forth.gr}}}
\address{$^1$Department of Mathematics and Applied Mathematics, University of Crete - Voutes Campus, 700 13 Heraklion, Greece.}
\address{$^2$Archimedes Center for Analysis, Modeling $\&$ Computation, University of Crete - Voutes Campus, 700 13 Heraklion, Greece.}
\address{$^3$Institute of Applied and Computational Mathematics, Foundation for Research and Technology, Nikolaou Plastira 100, Vassilika Vouton,
700 13 Heraklion, Greece.}
\date{today}

\begin{abstract}
We consider the singular semiclassical initial value problem for \textit{the phase space Schr\"{o}dinger equation}. We approximate semiclassical quantum evolution in phase space by analyzing initial states as superpositions of Gaussian wave packets and applying individually semiclassical anisotropic Gaussian wave packet dynamics, which is based on the \textit{the  nearby orbit approximation}; we accordingly construct a semiclassical approximation of the phase space propagator, \textit{the semiclassical wave packet propagator}. By the semiclassical propagator we construct asymptotic solutions of the phase space Schr\"{o}dinger equation, noting the connection of this construction to \textit{the initial value repsresentations}.
\end{abstract}

\section{Introduction}

Phase space formulations of quantum mechanics are used to describe microprocesses strongly influenced by their external environment, by representing mixed quantum states in terms of phase space quasiprobability distributions, which are used to express expectation values or classical energy densities and fluxes as phase space averages \cite{zachos}. Despite the increase in complexity of these formulations compared to the Schr\"{o}dinger representation (nonuniqueness of phase space quasidensities, doubling of variables and nonlocality of evolution equations), they prove worth studying even outside the context of the theory of open quantum systems, for a series of reasons. 

The phase space is the conceptually natural setting of quantum mechanics, more so in the setting of its explicit correspondence to classical mechanics, \textit{the semiclassical regime}. All inherent invariance properties of classical mechanics which become naturally manifest in phase space are reflections of approximate quantum mechanical symmetries, explicit in its phase space representations. Also, the problem of caustics (singularities void of physical content caused by projecting the dynamics to the configuration space) is resolved since they are not encountered as persistent obstacles toward global solutions in phase space formulations as they do in semiclassical considerations in configuration space. Finally, a unified approach to the two main classes of semiclassical states, WKBM (Wentzel-Kramers-Brillouin-Maslov) states and coherent states, becomes plaussible in phase space (e.g., \cite{hellerlitl}).

\textit{The wave packet representation} is a phase space representation intimately related \textit{to deformation quantization} \cite{zachos}, proposed in the works of Torres Vega and Frederick \cite{torres}, and also Harriman \cite{harriman}, Chruscinski and Moldawski \cite{polish}, de Gosson \cite{Gosson3} and Nazaikinskii \cite{naz2}, to mention a few; in its core there is the correspondence of pure classical states, i.e., single phase space points, to isotropic Gaussian wave packets,
$$G_{(q,p)}(x;\hbar)=(\pi\hbar)^{-d/4}e^{\frac{i}{\hbar}p\cdot (x-q)}e^{-\frac{1}{2\hbar}(x-q)^2} \ , \ \ (q,p)\in\mathbb{R}^{2d} \ ,$$
localized at that phase space point on the Heisenberg scale, $O(\sqrt{\hbar})$, so that $\langle\hat x\rangle=q$ and $\langle\hat p_x\rangle=p$. In this representation, \textit{the phase space wave function}, $\Psi(q,p)$, is defined as the coefficient of the isotropic Gaussian wave packet superposition of the wave function, at a given phase space point, 
$$\psi(x)=\Big(\frac{1}{2\pi\hbar}\Big)^{d/2}\int\Psi(q,p) \,G_{(q,p)}(x;\hbar)\, dqdp \ .$$
This coherent state resolution, \textit{the  wave packet transform}, also known as the Fourier-Bros-Iagolnitzer transform \cite{bargmann1,bargmann2,fefferman,naz2}, is closely related to \textit{the Bargmann transform}. The dynamics of the phase space wave function satisfies the nonlocal \textit{phase space Schr\"{o}dinger equation}. While the apparent drawback of this representation is that it cannot account for microscopic systems in strong interaction with their environment, it is a simpler representation than the Wigner formalism, which is quadratic in the wave function. 

In the wave packet representation, the phase space Schr\"{o}dinger spectral problem is well understood, e.g., the works of Luef and de Gosson \cite{Gosson1}, who have derived the eigenvalue equation departing from Moyal's spectral equation. However, understanding the corresponding \textit{initial value problem} is at the stage of infancy. The deeper understanding of the initial value problem, its asymptotic investigation toward a direct theory of semiclassical dynamics in phase space is a challenging and invaluable contribution to phase space quantum mechanics, unifying approaches taken from the field of Fourier integral operators \cite{hormander,FIO} to the theory of the Maslov canonical operator \cite{maslov1}, etc. Important contributions in this directions have been made by Oshmyan et al and Nazaikinksii et al \cite{naz2,naz1}.

A fundamentel semiclassical approximation of quantum dynamics is \textit{the Gaussian approximation}; the idea of approximating quantum evolution with the dynamics of \textit{an individual} Gaussian wave packet translated along an orbit is traced back to the foundational work of Schr\"{o}dinger \cite{schr}. Isotropic Gaussian wave packets provide natural semiclassical approximations of quantum states for free motion, as they are localized in phase space on the Heisenberg scale $O(\sqrt{\hbar})$, `occupying' a Planck cell centered at $(q,p)$, and exhibiting oscillations at the de Broglie wavelength $O(\hbar)$ (for a general account on coherent states, see \cite{cohstates}). 

Heller et al \cite{Huber} argued for the use of both single isotropic and anisotropic Gaussian wave packets as approximations to the propagation of initially Gaussian wave packets, based on \textit{the  nearby orbit approximation}, while in \cite{hellerlitl}, Huber, Heller and Littlejohn showed that isotropic Gaussian wave packet dynamics can stand as a generalization of complex phase WKBM semiclassical propagation. In \cite{Hagedorn}, Hagedorn showed that the an initially Gaussian state retains its Gaussian form within a certain timescale under quantum dynamics.

These ideas have been implemented in the solution of the more general problem of establishing an asymptotic solution of the Schr\"{o}dinger equation or the wave equation along a given curve, modulated by a Gaussian profile, known as \textit{Gaussian beams}. Gaussian beams find applications in a range of physical problems, in optics, accoustics, etc (see, e.g., \cite{riccati beams,chinese}). In the case of anisotropic Gaussian wave packets, the ansatz
$$G^Z_{(q,p)}(x;\hbar)=(\pi\hbar)^{-d/4}(\det\, Z_2)^{1/4}e^{i\phi}e^{\frac{i}{\hbar}p\cdot (x-q)}e^{\frac{i}{2\hbar}(x-q)\cdot Z(x-q)} \ , \ \ (q,p)\in\mathbb{R}^{2d} \ ,$$ 
for some real phase $\phi$, the anisotropy quadratic form, which satisfies $Z_2:={\rm Im}\, Z\succ 0$, is shown to obey the dynamics of the initially nearby orbit approximation. These dynamics are equivalent to a matrix Riccati initial value problem for the anisotropy matrix $Z$, common to all Gaussian beam asymptotic solutions \cite{Keller,riccati beams,chinese,initial data representation}. In \cite{litt1}, Littlejohn generalized the works of Heller et al. on the thawed, or anisotropic, Gaussian dynamics for generic initially localized states, by constructing a nonlinear quantum propagator, as an explicit composition of Weyl and metaplectic operators. Closely related to Heller's and Littlejohn's works, Maslov and Shvedov \cite{maslov QFT1, maslov QFT2, maslov QFT3} suggested the dynamics of anisotropic Gaussian wave packets, giving an alternative representation of the dynamics of the anisotropy quadratic form, called by some authors \textit{the  Maslov-Shvedov complex germ}, as a one parameter flow in the Siegel upper half space, related to the Weil representation of the metaplectic group therein.

The Gaussian approximation, however, breaks down as localization is lost in an irreversible spreading at a certain timescale, an effect suppressed only for quadratic potentials. Added to the above, the fact that for quadratic potentials the evolution of coherent \textit{superpositions} of phase space eigenfunctions result in expressions reminiscent of the evolution of single Gaussian wave packets, hints the method of approximating the evolution of a quantum state \textit{by a superposition of semiclassically propagated Gaussian wave packets}. The starting point of this approximation is the resolution of the identity in quantum state space in the overcomplete basis of coherent states, in particular isotropic Gaussian wave packets; 
$$U^t=\Big(\frac{1}{2\pi\hbar}\Big)^{d}\int U^t G_{(q,p)} \langle G_{(q,p)},\cdot \rangle_{L^2(\mathbb{R}^d)} \,dqdp \ .$$
One makes an explicit ansatz for the propagation of a single wave packet under the Schr\"{o}dinger propagator \cite{naz1,rober1}, $U^t G_{(q,p)}$. The totality of approximating for the single Gaussian dynamics amount to \textit{the initial value representations} of quantum dynamics. The first work in this direction was that of Herman and Kluk \cite{HK}, who argued on the validity of approximating semiclassical evolution by analyzing wave functions by a multitude of nonspreading isotropic Gaussian wave packets, their shape held rigid, modulated by some overall amplitude and phase factor, begining from the van Vleck-Gutzwiller approximation of the propagator. More recently, Rousse and Robert \cite{rousse,rober HK} setting off from a `head down' direction, assumed a semiclassical time evolution for generic initial data of the Schr\"{o}dinger equation, in terms of a Fourier integral operator identified with the Herman-Kluk propagator, in order to justify this approximation on the basis of rigorous estimates for the asymptotic solutions.

In this paper we turn attention to semiclassical asymptotic solutions of the initial value problem of the phase space Schr\"{o}dinger equation, rather than giving phase space representations of solutions of the Schr\"{o}dinger equation. Having defined the phase space quantum flow and phase space Schr\"{o}dinger propagator, we construct a semiclassical propagator, based on the anisotropic Gaussian approximation, which is closely related to the nearby orbit approximation. As the anisotropic Gaussian approximation is made for the totality of orbits, all of which are taken into account in the wave packet resolution of the phase space wave function, the semiclassical propagator admits generic semiclassical initial data, not just localized ones. Remarking on the underlying metaplectic structure of the semiclassical phase space propagator, we construct asymptotic solutions of the phase space Schr\"{o}dinger for phase space WKBM initial data, as a half density semiclassically localized on the complex almost analytic extension of the Lagrangian manifold of the solution of the corresponding Hamilton-Jacobi equation.

We consider Hamiltonian systems and their quantization in configuration space $\mathbb{R}^d$, with Cartesian coordinates $x=(x_1,\ldots,x_d)$. As a trivial vector bundle, the phase space $\mathbb{R}^d\oplus \mathbb{R}_d\cong \mathbb{R}^{2d}$, has the structure of a linear symplectic space with canonical coordinates $(q,p)=(q_1,\ldots,q_d,p_1,\ldots,p_d)$ and complex pseudocoordinates\footnote{The transformation $(q,p)\mapsto (z,\bar z)$ would be a symplectic pseudocoordinate transformation for the normalization $(z,\bar z)=(\frac{q-ip}{\sqrt{2}},\frac{q+ip}{\sqrt{2}})$, so that $|\frac{\partial(z,\bar z)}{\partial (q,p)}|=1$.} $(z,\bar z)=(q-ip,q+ip)$, where $z\in\mathbb{C}^d$. For the action of the Hamiltonian flow $g^t:\mathbb{R}^{2d}\rightarrow \mathbb{R}^{2d}$, we use the notation $(q_t,p_t)=(\underline q(t;q,p),\underline p(t;q,p)):=g^t(q,p)$ for the terminal point of the orbit $\gamma^t(q,p)$ eminating from $(q,p)$. As the flow is taken to be generated by an autonomous Hamiltonian, generic apart from smoothness assumptions, including chaotic Hamiltonians, we assume short time semiclassical propagation, $t=o\Big(\log\hbar\Big)$, within \textit{the Ehrenfest time} (e.g., see \cite{heller1}).

\section{The Phase Space Schr\"{o}dinger Equation}

\subsection{The Wave Packet Transform}
\textit{The wave packet representation} is an equivalent representation of quantum mechanics set in phase space, as the position representation is set in configuration space \cite{torres}. By the Stone-von Neumann theorem \cite{JvN}, the two representations are related by means of a linear, unitary map, \textit{the wave packet transform}\footnote{All linear representations of the Schr\"{o}dinger position realization of quantum mechanics in $L^2(\mathbb{R}^d)$ into a phase space realization in $L^2(\mathbb{R}^{2d})$, are equivalent modulo phase \cite{JvN}. This can be verified by the uniqueness, up to phase, of the kernel of the wave packet transform \cite{bargmann1,bargmann2}.} \cite{torres,naz2,naz1}, which maps configuration space wave functions $\psi(x)$ to phase space wave functions $\Psi(q,p)$,
\begin{equation}
\mathcal{W}:L^2(\mathbb{R}^d,\mathbb{C};dx)\rightarrow L^2(\mathbb{R}^{2d},\mathbb{C};dqdp) \ .
\end{equation}
It is given explicitly by the integral transform \cite{torres,naz2,naz1}
\begin{equation}
\Psi(q,p)=\mathcal{W}\psi(q,p)=\Big(\frac{1}{2\pi\hbar}\Big)^{d/2}\int\bar G_{(q,p)}(x;\hbar)\psi(x)\,dx \ .
\label{eq:wpt}
\end{equation}
The kernel of the transform is the isotropic Gaussian wave packet with base point $(q,p)\in\mathbb{R}^{2d}$, complex conjugated,
\begin{equation}
G_{(q,p)}(x;\hbar)=(\pi\hbar)^{-d/4}e^{\frac{i}{\hbar}\Big(p\cdot (x-q)+\frac{i}{2}(x-q)^2\Big)} \ .
\label{eq:g}
\end{equation}

The wave packet transform is an isometry, in the sense that if $\psi \in L^2(\mathbb{R}^d)$ is normalized to unity, then
\begin{equation}
\int|\mathcal{W} \psi(q,p)|^2\,dqdp=\int|\psi(x)|^2\,dx=1 \ ,
\end{equation}
while $\|G_{(q,p)}\|_{L^2(\mathbb{R}^d)}=1$.

As a linear map between Hilbert spaces, the wave packet transform is not `onto'; its image is a subspace of $L^2(\mathbb{R}^{2d})$, \textit{the Fock-Bargmann space} (see ($\ref{eq:ker1}$) of the appendix), 
\begin{equation}
\mathfrak{F}^2:= {\rm im}\,(\mathcal{W})\subset L^2(\mathbb{R}^{2d}) \ ,
\label{eq:fockspace}
\end{equation}
characterized by \textit{the Fock-Bargmann constraint} \cite{torres,harriman,polish,naz1,naz2}
\begin{equation} 
\Big(\hbar\frac{\partial}{\partial q}-i\hbar\frac{\partial}{\partial p}-ip \Big)\Psi=0 \ ,
\end{equation}
or
\begin{equation}
\Big(\frac{\partial}{\partial q}-i\frac{\partial}{\partial p} \Big)\Big(e^{p^2/2\hbar}\Psi\Big)=0 \ ,
\label{eq:twist}
\end{equation}
implying that \textit{not all square integrable phase space functions constitute phase space wave functions} \cite{bargmann1,torres,naz2,naz1}. The Fock-Bargmann space $\mathfrak{F}^2$ possesses special analyticity properties; it is essentially the space of Gaussian weighted square integrable analytic functions (see subsection $(\ref{analytic})$ of the appendix). 

The wave packet transform microlocalizes quantum states in phase space; the real part of the phase of the kernel contributes to semiclassical localization in the momenta, while the imaginary part to semiclassical localization in the positions.

For $\Psi\in\mathfrak{F}^2$, the inverse wave packet transform is defined as \cite{naz2,naz1} 
\begin{equation}\label{eq:iwpt}
\psi(x)=\mathcal{W} ^{-1}\Psi(x)=\Big(\frac{1}{2\pi\hbar}\Big)^{d/2}\int G_{(q,p)}(x;\hbar)\Psi(q,p)\,dqdp \ .
\end{equation}

\subsection{The Phase Space Schr\"{o}dinger Equation}
We consider the semiclassical Schr\"{o}dinger initial value problem corresponding to a mechanical system described by the Hamiltonian function $H$,
\begin{eqnarray}
\hspace{-20mm}i\hbar\frac{\partial \psi}{\partial t}=\hat H\psi=H\Big(\stackrel{2}{x},-i\hbar\stackrel{1}{\partial}_x\Big)\psi \ , \ \ 0<t\leqslant T \nonumber \\
\hspace{-20mm}\psi( \cdot ,0)=\psi_0^\hbar\in L^2(\mathbb{R}^d) \ ,
\label{eq:ivp}
\end{eqnarray}
the Feynman indices stating the order of action of the symbol noncommuting operator arguments. The dependence of the initial data on $\hbar \in\,]0,1]$ is assumed essentially singular at $\hbar=0$. For the standard form, $H(q,p)=p^2+V(q)$, the above problem takes the standard Schr\"{o}dinger form, for which operational ordering is unambiguous
\begin{eqnarray}
\hspace{-20mm}i\hbar\frac{\partial \psi}{\partial t}=-\hbar^2\Delta_x\psi+V\psi \ , \ \ 0<t\leqslant T \nonumber \\
\hspace{-20mm}\psi( \cdot ,0)=\psi_0^\hbar\in L^2(\mathbb{R}^d) \ ,
\end{eqnarray}
where $-\Delta_x=-\sum_j\partial^2_{x_j}$ is the positive Laplacian in $\mathbb{R}^d$.

We consider semiclassical time evolution in the above problem for short times, which is quantified as $T=o\Big(\log \hbar\Big)$, i.e., within the Ehrenfest time scale. This is assumed in order to ensure the validity of the semiclassical approximation in the case the underlying Hamiltonian flow is chaotic \cite{heller1}.

The phase space representation of the Heisenberg quantized Hamiltonian, $\hat H$, is (see \cite{torres,naz2,naz1} and $(\ref{eq:Toplitz})$ of the appendix)
\begin{equation}
\hspace{-20mm}\check H=\mathcal{W}\circ \hat H\circ \mathcal{W}^{-1}=\mathcal{W}\circ H\Big(\stackrel{2}{x},-i\hbar\stackrel{1}{\partial}_x\Big)\circ \mathcal{W}^{-1}=H\Big(\stackrel{2}{q}+i\hbar\partial_p,-i\hbar\stackrel{1}{\partial}_q\Big) \ .
\end{equation} 

Therefore, the wave packet transform of the  initial value problem ($\ref{eq:ivp}$), the  initial value problem for \textit{the phase space Schr\"{o}dinger equation} reads
\begin{eqnarray}
\hspace{-20mm}i\hbar\frac{\partial \Psi}{\partial t}=\check H\Psi=H\Big(\stackrel{2}{q}+i\hbar\partial_p,-i\hbar\stackrel{1}{\partial}_q\Big)\Psi \ , \ \ 0<t\leqslant T\nonumber \\
\hspace{-20mm}\Psi( \cdot ,0)=\Psi^\hbar_0:=\mathcal{W}\psi^\hbar_0\in\mathfrak{F}^2 \ ,
\label{eq:pivp}
\end{eqnarray}
for the phase space wave function $\Psi=\mathcal{W}\psi$. For for the standard form 
\begin{eqnarray}
\hspace{-20mm}i\hbar\frac{\partial \Psi}{\partial t}=-\hbar^2\Delta_q\Psi+V\Big(q+i\hbar\frac{\partial}{\partial p}\Big)\Psi \ , \ \ 0<t\leqslant T \nonumber \\
\hspace{-20mm}\Psi( \cdot ,0)=\Psi^\hbar_0:=\mathcal{W}\psi^\hbar_0\in\mathfrak{F}^2 \ ,
\end{eqnarray}
with $-\Delta_q=-\sum_j\partial_{q_j}^2$ the positive position Laplacian in the phase space. 

The wave packet transform retains the essential singularity of the equation itself as well as of the initial data, as will be shown in section (\ref{solutions}) for WKBM states as prototype semiclassical states. This renders the phase space image of the initial problem, in complete analogy, semiclassically singular.

A qualitative difference from the Schr\"{o}dinger equation in the case of the standard Hamiltonian form arises from the potential term; for nonpolynomial potentials, this term is a pseudodifferential operator, which introduces nonlocality as well as transport effects in phase space dynamics, both common features in phase space evolution equations, such as for the von Neumann equation. 

In the case the potential is real analytic, this pseudodifferential operator term is defined by the Taylor series
\begin{equation}
V(q+i\hbar \partial_p)\Psi^\hbar (q,p)=\sum_{\alpha\in\mathbb{Z}^d}\frac{(i\hbar)^{|\alpha|}}{\alpha !}\partial^{\alpha}_qV(q)\partial _p^{\alpha}\Psi^\hbar(q,p) \ ,
\end{equation}
while in general, we have the representation
\begin{equation}
\hspace{-20mm}V\Big(q+i\hbar\frac{\partial}{\partial p}\Big)\Psi^\hbar(q,p)=\Big(\frac{1}{2\pi\hbar}\Big)^{d}\int e^{\frac{i}{\hbar}\eta\cdot (\xi-p)}V(q+\eta)\Psi^\hbar(q,\xi)\,d\eta d\xi \ .
\end{equation}

\subsection{Dynamics of the Phase Space Wave Function}

The solution of the Schr\"{o}dinger initial value problem ($\ref{eq:ivp}$), with generic initial data $\psi_0^\hbar\in L^2(\mathbb{R}^{d})$, is given by the action of the Schr\"{o}dinger time evolution operator (quantum flow) on the initial data,
\begin{equation}
\psi^\hbar(x,t)= U^t\psi^\hbar_0(x)=e^{-\frac{i}{\hbar}t\hat H}\psi^\hbar_0(x)=\int K(x,y,t;\hbar)\psi^\hbar_0(y)\,dy \ ,
\label{eq:Kprop}
\end{equation}
$K$ being \textit{the Schr\"{o}dinger propagator}, the Schwartz kernel of $U^t$.

By beginning with the completeness relation of the isotropic Gaussian wave packets, a resolution of the identity in $L^2(\mathbb{R}^d)$ \cite{rober1}, 
\begin{equation}
\Big(\frac{1}{2\pi\hbar}\Big)^{d}\int\bar G_{(q,p)}(x;\hbar)  G_{(q,p)}(y;\hbar)\,dqdp=\delta(x-y) \ , 
\label{eq:completeness}
\end{equation}
we acquire a phase space resolution of the Schr\"{o}dinger propagator
\begin{equation}
\hspace{-20mm}K(x,y,t;\hbar)=\Big(\frac{1}{2\pi\hbar}\Big)^{d}\int U^tG_{(q,p)}(x;\hbar)\bar G_{(q,p)}(y;\hbar)\,dqdp \ .
\end{equation}

By applying the wave packet transform ($\ref{eq:wpt}$) on the representation formula ($\ref{eq:Kprop}$), we derive \textit{the phase space quantum flow}, $\mathcal{U}^t$, which yields the phase space wave function
\begin{equation}
\hspace{-20mm}\Psi^\hbar(q,p,t)=\mathcal{U}^t\Psi^\hbar_0(q,p)=e^{-\frac{i}{\hbar}t\check H}\Psi_0^\hbar(q,p)=\int \mathcal{K}(q,p,\eta,\xi,t;\hbar)\Psi^\hbar_0(\eta,\xi)\,d\eta d\xi \ ,
\label{eq:pswf}
\end{equation}
$\mathcal{K}$ being \textit{the phase space Schr\"{o}dinger propagator}, the Schwartz kernel of $\mathcal{U}^t$. The phase space Schr\"{o}dinger propagator $\mathcal{K}$ is expressed in terms of the Schr\"{o}dinger propagator, $K$, by
\begin{eqnarray}
\hspace{-20mm}\mathcal{K}(q,p,\eta,\xi,t;\hbar)=\Big(\frac{1}{2\pi\hbar}\Big)^{d} \int \!\!\!\int\bar G_{(q,p)}(x;\hbar)G_{(\eta,\xi)}(y;\hbar)K(x,y,t;\hbar)\,dxdy\nonumber \\
\hspace{-20mm}=\Big(\frac{1}{2\pi\hbar}\Big)^{d} \int \bar G_{(q,p)}(x;\hbar)U^tG_{(\eta,\xi)}(x;\hbar)\,dx  \ .
\label{eq:psprop}
\end{eqnarray}
It can be easily checked that ($\ref{eq:pswf}$) solves the problem ($\ref{eq:pivp}$), for generic initial data $\Psi_0^\hbar\in\mathfrak{F}^2$.

The above representation,
\begin{equation}
\mathcal{K}(q,p,\eta,\xi,t;\hbar)=\Big(\frac{1}{2\pi\hbar}\Big)^{d/2}\mathcal{W}\circ U^t G_{(\eta,\xi)}(q,p;\hbar) \ ,
\label{eq:KG}
\end{equation}
serves as a starting point for \textit{Gaussian approximations} for phase space quantum dynamics; as we shall see subsequently, these ammount to making a specific approximation for the dynamics of the single Gaussian wave packet, $U^t G_{(q,p)}(x;\hbar)$; the phase space propagator is, then, proportional to the wave packet transform of the evolved Gaussian wave packet.

It follows that like time evolution in configuration space, time evolution in phase space is also generated by a unitary group 
\begin{equation}
\mathcal{U}^t=\mathcal{W}\circ U^t\circ  \mathcal{W} ^{-1}=e^{-\frac{i}{\hbar}t\check H} \ \ {\rm on} \ \ \mathfrak{F}^2 \ ,
\label{eq:U}
\end{equation}
constituting the phase space propagator $\mathcal{U}^t$ a unitary quantum flow, in the sense that, on $\mathfrak{F}^2$, 
\begin{equation}
\mathcal{U}^{t \, *}=\mathcal{U}^{-t} \ , \ \ \mathcal{U}^t\circ\mathcal{U}^{t'}=\mathcal{U}^{t+t'} \ , \ \ \mathcal{U}^0={\rm Id}_{\mathfrak{F}^2} \ , \ \ t,t'\in\mathbb{R} \ ,
\end{equation}
so that phase space dynamics induced by $\mathcal{U}^t$ retain the Gaussian `twisted' analyticity of the initial data (see relations $(\ref{eq:ta1})$ and $(\ref{eq:ta6})$ of the appendix). 

As only those $\Psi\in\mathfrak {F}^2$ correspond to Schr\"{o}dinger wave functions on configuration space, only `twisted' analytic phase space functions correspond to \textit{pure quantum states}, while all other choices of $L^2$ phase space functions correspond to \textit{mixed quantum states}.

In terms of the phase space kernel, unitarity becomes manifest by the following
\begin{equation}
\hspace{-20mm}\int\mathcal{K}(q,p,u,v,t;\hbar)\mathcal{K}(u,v,\eta,\xi, t';\hbar)\,dudv=\mathcal{K}(q,p,\eta,\xi,t+ t';\hbar) \ ,
\end{equation}
and $\bar\mathcal{K}(\eta,\xi,q,p,-t;\hbar)=\mathcal{K}(q,p,\eta,\xi,t;\hbar)$.

The Schr\"{o}dinger propagator $K$ is a weak solution of the Schr\"{o}dinger equation,
\begin{equation}
i\hbar\frac{\partial K}{\partial t}(x,y,t;\hbar)-\hat H_xK(x,y,t;\hbar)=i\hbar\delta(t)\delta(x-y) \  .
\end{equation}
Taking the wave packet transform of the above, we have the analogous equation for the phase space propagator
\begin{equation}
\hspace{-20mm}i\hbar\frac{\partial \mathcal{K}}{\partial t}(q,p,\eta,\xi,t;\hbar)-\check H_{(q,p)}\mathcal{K}(q,p,\eta,\xi,t;\hbar)=i\hbar\delta(t)\langle G_{(q,p)},G_{(\eta,\xi)}\rangle_{L^2(\mathbb{R}^d)} \  .
\end{equation}

Due to the Gaussian integration in ($\ref{eq:psprop}$), the kernel $\mathcal{K}$ has stronger smoothness properties than $K$. Firstly, it has well defined values across the hyperplane $\Big\{(q,p,\eta,\xi):\,(q,p)=(\eta,\xi)\Big\}$, while for $t\rightarrow 0^+$ it does not converge weakly to a Dirac distribution, but rather is a Dirac mollifier on the Heisenberg scale (see ($\ref{eq:Berg1}$)-($\ref{eq:Berg3}$) of the appendix),
\begin{equation} 
\hspace{-20mm}\lim_{t\rightarrow 0^+}\mathcal{K}(q,p,\eta,\xi,t;\hbar)=\Big(\frac{1}{2\pi\hbar}\Big)^{d} \int \bar G_{(q,p)}(x;\hbar)G_{(\eta,\xi)}(x;\hbar)\,dx \ ,
\end{equation}
which we define as $\mathcal{K}(q,p,\eta,\xi,0;\hbar)$.

As an operator acting in $L^2(\mathbb{R}^{2d})$, the initial phase space propagator is the kernel of the projector (see ($\ref{eq:projector1}),(\ref{eq:projector2}$) of the appendix) onto the subspace $\mathfrak{F}^2$,
\begin{equation}
\mathcal{P}_{\mathfrak{F}^2}=\mathcal{W}\circ \mathcal{W}^*:\, L^2(\mathbb{R}^{2d})\rightarrow \mathfrak{F}^2 \ ,
\end{equation}
meaning that if $\Psi^\hbar_0\in\mathfrak{F}^2$, then
\begin{equation}
\hspace{-20mm}\mathcal{U}^0\Psi_0^\hbar(q,p)=\int\mathcal{K}(q,p,\eta,\xi,0;\hbar)\Psi_0^\hbar(\eta,\xi)\,d\eta d\xi=\mathcal{P}_{\mathfrak{F}^2}\Psi_0^\hbar(q,p)=\Psi_0^\hbar(q,p) \ ,
\end{equation}
so that for $t=0$ the representation of the solution $\Psi^\hbar(q,p,t)$ reconstructs the initial data, $\Psi_0$, 
\begin{eqnarray}
\hspace{-20mm}\Psi^\hbar(q,p,0)=\Big(\frac{1}{2\pi\hbar}\Big)^{d} \int \!\!\!\int\bar G_{(q,p)}(x;\hbar)G _{(\eta,\xi)}(x;\hbar)\Psi_0^\hbar (\eta,\xi)\,dxd\eta d\xi \nonumber \\
\hspace{-20mm}=\Big(\frac{1}{2\pi\hbar}\Big)^{3d/2}\int\,dy\,\psi_0^\hbar (y)\int \!\!\!\int\bar G_{(q,p)}(x;\hbar)G _{(\eta,\xi)}(x;\hbar)\bar G _{(\eta,\xi)}(y;\hbar)\, dxd\eta d\xi \nonumber \\
\hspace{-20mm}=\Big(\frac{1}{2\pi\hbar}\Big)^{d/2}\int\bar G _{(q,p)}(x;\hbar)\psi_0^\hbar (x)\,dx=\Psi ^\hbar_0(q,p) \ , 
\end{eqnarray}
by utilizing the completeness property. 

This reproducing property of the initial propagator is a result of the underlying analytic structure of the Fock-Bargmann space $\mathfrak{F}^2$, something which is elaborated on in the appendix (subsection $(\ref{analytic})$ of the appendix).

\section{Anisotropic Wave Packet Dynamics}

In this section we deal with the propagation of Gaussian wave packets in configuration space and their phase space image under the wave packet transform.  In particular, we consider \textit{the anisotropic Gaussian wave packet}, following the independent works of Heller, Huber, Littlejohn (isotropic case) \cite{Huber,hellerlitl} and Littlejohn \cite{litt1}, Maslov, Shvedov \cite{maslov QFT1,maslov QFT2} (anisotropic case) and more recently Robert \cite{rober1}, Nazaikinskii, Schulze, Sternin \cite{rober1,naz1}, and Stoyanovsky \cite{stoyanovsky}. An excellent account of the subject, masterfully touching upon its dynamical and algebraic aspects, is given by Littlejohn \cite{litt1}.

As we shall see, the semiclassical dynamics of anisotropic Gaussian wave packets in configuration space realizes a representation of the metaplectic group on the Siegel upper half space, and is equivalent to the variational system of the Hamiltonian flow in the frame of \textit{the nearby orbit method} (see, e.g., \cite{stoyanovsky}).

\subsection{Anisotropic Wave Packet Dynamics in Configuration Space}

An attempt for the semiclassical description of the dynamics in the position representation has come under the name of \textit{initial value representations}, involving superpositions of initial semiclassical states in wave packets, $G_{(q,p)}(x;\hbar)$, propagating them in a certain approximation along the Hamiltonian orbits $\gamma^t(q,p)$ eminating from $(q,p)$ (see introduction), and then superposing the evolved wave packets back together, with respect to \textit{all} initial phase space points $(q,p)$.

Here we consider \textit{the  anisotropic Gaussian wave packet approximation}, in order to construct a semiclassical propagator for the phase space Schr\"{o}dinger equation. The leading term in the semiclassical propagation of a single normalized isotropic Gaussian wave packet, amounts to the anisotropic Gaussian approximation, namely $U^tG_{(q,p)}(x;\hbar)$ retaining its Gaussian form, yet acquiring an anisotropic phase. In particular, in this approximation, the initial state
\begin{equation}
G_{(q,p)}(x;\hbar) =(\pi\hbar)^{-d/4}e^{\frac{i}{\hbar}\Big(p\cdot (x-q)+\frac{i}{2}(x-q)^2\Big)} \ ,
\end{equation}
is evolved semiclassically to the anisotropic Gaussian wave packet moving along the ray $x=q_t$, and semiclassically concentrated on that point on the Heisenberg scale \cite{naz1,stoyanovsky}. In particular, we consider the following ansatz
\begin{equation}
\hspace{-20mm}G^Z_{(q,p)}(x,t;\hbar)=u(q,p,t;\hbar) e^{\frac{i}{\hbar}\Big(A(q,p,t)+p_t\cdot (x-q_t)+\frac{1}{2}(x-q_t)\cdot Z(q,p,t)(x-q_t)\Big)} \ ,
\label{eq:MSwp}
\end{equation}
where $A$ is a real valued phase, and the matrix $Z=Z_1+iZ_2$ is symmetric and has positive definite imaginary part, $Z^T=Z$ and $Z_2\succ 0$, to be determined by the dynamics.

By substituting the ansatz ($\ref{eq:MSwp}$) in the Schr\"{o}dinger equation, and demanding that it be an asymptotic solution, in the sense that for a certain time scale and uniformly in $x$,
\begin{equation}
\Big(i\hbar\frac{\partial}{\partial t}-\hat H\Big)G^Z_{(q,p)}(x,t;\hbar)=O(\hbar) \ ,
\label{eq:sclsol}
\end{equation} 
we arrive at a system of characteristic differential equations along the orbit $\gamma^t(q,p)$ \cite{litt1,rober1,naz1,stoyanovsky}, analogous to the WKBM system; the zeroth order equation in $\sqrt{\hbar}$ is 
\begin{equation}
\frac{\partial A}{\partial t}(q,p,t)=L(q,p,t) \ ,
\end{equation}
along with the initial condition $A(q,p,0)=0$, where $L$ is the phase space Lagrangian of the flow. These give the phase space action along the orbit $\gamma^t(q,p)$,
\begin{equation}
A(q,p,t)=\int_0^tp_\tau\cdot \dot q_\tau \,d\tau-H(q,p)t \ .
\label{eq:action}
\end{equation}
The following properties of the phase space action shall become of use in what follows,
\begin{equation}
\frac{\partial A}{\partial q}=-p+p_t\cdot \frac{\partial q_t}{\partial q} \ , \ \ \frac{\partial A}{\partial p}=p_t\cdot \frac{\partial q_t}{\partial p} \ ,
\label{eq:dA}
\end{equation} 
where appropriate matrix multiplication and contraction is implied.

The first order equation in $\sqrt{\hbar}$ yields Hamilton's equations for $(q_t,p_t)$, which are automatically statisfied along the orbit $\gamma^t(q,p)$. The second order equation gives the amplitude $u$ and the anisotropy form $Z$. The amplitude is 
\begin{eqnarray}
\hspace{-20mm}u(q,p,t;\hbar)=(\pi\hbar )^{-d/4}\Big({\rm det}\,{\rm Im}\, Z(q,p,t)\Big)^{1/4}e^{i\phi(q,p,t)} \nonumber \\
=(\pi\hbar )^{-d/4}\exp\Big(-\frac{1}{2}\int_0^t{\rm tr}\, H''_{pp}Z(q,p,\tau)\,d\tau\Big) \ ,
\label{eq:u}
\end{eqnarray}
offering an additional phase, or index to be accurate,
\begin{equation}
\phi(q,p,t)=-\frac{1}{2}\int_0^t{\rm tr}\,H_{pp}''Z_2(q,p,\tau)\, d\tau \ .
\end{equation}

The dynamics of the anisotropy form, sometimes called \textit{the  Maslov-Shvedov complex germ}, \textit{the matrix Riccati equation} \cite{hellerlitl,riccati,rober1,naz1,stoyanovsky}, essentially controling the direction, the shape and the spreading of the propagated state, 
\begin{equation}
\dot Z+ZH''_{pp}Z+H''_{pq}Z+ZH''_{qp}+H_{qq}''=0 \ ,
\label{eq:riccati}
\end{equation}
with the initial condition $Z(0)=iI$.

The time dependence of the anisotropy form is implicit in the flow $(q,p)\mapsto (q_t,p_t)$, as the matrix Riccati dynamics is taken along the orbit $\gamma^t(q,p)$; the anisotropy form $Z$ depends on time only through $(q_t,p_t)$, except for the class of quadratic Hamiltonians which yield give solutions $Z(t)$ independent of the initial point $(q,p)$. We thus make the notational convention for the functional dependence $Z(q,p,t):=Z(q_t,p_t)$, always for nonquadratic potentials. 

The wave packet character of $G^Z$ is guaranteed, as by beginning at $Z(0)=iI$, $Z$ remains symmetric with positive definite imaginary part for all times. We stress that \textit{the Riccati equation guarantees unitarity of the semiclassical flow $G_{(q,p)}\mapsto G^Z_{(q,p)}$}, in the sense that $\|G^Z_{(q,p)}(\cdot\, ,t)\|_{L^2(\mathbb{R}^d)}=1$, for all times $t\geqslant 0$. As by assumption $G^Z\in L^2(\mathbb{R}^d)$, we must have $Z(t)^T=Z(t)$ and $Z_2(t)\succ 0$, the Riccati equation must define a flow in \textit{the  Siegel upper half space}
\begin{equation}
\mathbb{H}_d=\Big\{Z\in\mathbb{C}^{d\times d}:\, Z^T=Z \ , \ \ {\rm Im}\,Z\succ 0\Big\} \ ,
\end{equation}
which is a $d(d+1)$-dimensional manifold generalizing the complex upper half plane $\mathbb{H}=:\mathbb{H}_1=\{z\in\mathbb{C}:\,{\rm Im}\,z>0\}$ \cite{cartan,stoyanovsky}.

The anisotropic Gaussian wave packet is a semiclassical solution of the Schr\"{o}dinger equation for short times, $t\ll\log\hbar$, within the Ehrenfest time scale, with initial data
\begin{equation}
G^Z_{(q,p)}(x,0;\hbar)=G_{(q,p)}(x;\hbar) \ .
\end{equation}

The short time scale is required so that there is no sufficient spreading, sustaining localization on the Heisenberg scale is not lost. By semiclassical solution relation ($\ref{eq:sclsol}$) is meant, while there exists some $C(q,p,T)>0$ such that \cite{rober1}
\begin{equation}
\|U^tG_{(q,p)}- G^Z _{(q,p)}( \cdot ,t)\|_{L^2(\mathbb{R}^d)}\leqslant C(q,p,T)t\, \hbar^{1/2}
\end{equation}
and 
\begin{equation}
\sup_{0< t\leqslant T}\|U^tG_{(q,p)}- G^Z _{(q,p)}( \cdot ,t)\|_{L^2(\mathbb{R}^d)}\leqslant C(q,p,T)\hbar^{3/2} \ .
\end{equation}

\subsection{The Nearby Orbit Approximation and the Matrix Riccati Equation}

As noted in \cite{hellerlitl}, in anisotropic Gaussian wave packet dynamics the spatio-temporal variation in its localization is dictated by a certain matrix Riccati dynamics, which in turn is related to the separation dynamics of initially nearby Hamiltonian orbits. These dynamics are essentially common to all Gaussian narrow beam dynamics, i.e., dynamics for asymptotic solutions concentrated with a Gaussian profile about the characteristics \cite{riccati beams}.

The Riccati equation of the anisotropy form is equivalent to the linear variational system governing the stability of the Hamiltonian flow \cite{hellerlitl,rober1,naz1,stoyanovsky}. The Hamiltonian flow is given by the system
\begin{eqnarray}
\frac{dq_t}{dt}=\frac{\partial H}{\partial p}(q_t,p_t)\nonumber \\
\frac{dp_t}{dt}=-\frac{\partial H}{\partial q}(q_t,p_t) \ ,
\end{eqnarray}
with initial data $(q,p)$

The stability of initially nearby orbits is given by the linearization of the dynamics \textit{about a reference orbit}, rather than a fixed phase space point. We note that the deviation coordinates from a given propagated point $g^t(q_0,p_0)$, say $g^t(q_0,p_0)+(q_t,p_t)$ are canonical coordinate (retaining the same notation for the original coordinates and the `deviation' coordinates), so that the nonautonomous linearized flow for the deviation, is itself Hamiltonian 
\begin{equation}
\frac{d}{dt}
\left(\begin{array}{ccc}
q_t  \\
p_t   
\end{array}\right)=JH''(g^t(q_0,p_0))\left(\begin{array}{ccc}
q_t  \\
p_t   
\end{array}\right) \ , 
\end{equation}
where $J=\left(\begin{array}{ccc}
0 & I \\
-I & 0  
\end{array}\right)$ is the $2d\times 2d$ unit symplectic matrix.

The corresponding linear flow is given by a matrix symplectic transformation 
\begin{equation}
g_0^t=e^{t JH''}=\exp t \left(
\begin{array}{ccc}
H''_{pq} & H''_{pp} \\
-H''_{qp} & - H''_{qq}
\end{array}\right) \in {\rm Sp}(2d,\mathbb{R})\ ,
\end{equation}
evaluated along the orbit $\gamma^t(q,p)$. The \textit{variational system}, which relates the sensitivity of the system with respect to nearby initial points, is obtained by varying Hamilton's equations by the initial points $(q,p)$,
\begin{equation}
\frac{d}{dt}\frac{\partial}{\partial z}\left(\begin{array}{ccc}
q_t  \\
p_t   
\end{array}\right)=JH''(g^t(q,p))\frac{\partial }{\partial z}\left(\begin{array}{ccc}
q_t  \\
p_t   
\end{array}\right) \ ,
\end{equation}
where $z=q-ip$ is the pseudocomplexified phase space pseudocoordinate (see introduction).

Following, e.g. \cite{naz1}, we write the above as 
\begin{equation}
\frac{d}{dt}\left(
\begin{array}{ccc}
X \\
Y 
\end{array}\right)=\left(\begin{array}{ccc}
H''_{pq} & H''_{pp} \\
-H''_{qp} & -H''_{qq}  
\end{array}\right)\left(
\begin{array}{ccc}
X \\
Y 
\end{array}\right) \ ,
\label{eq:varsys}
\end{equation}
where 
\begin{eqnarray}
X=2\frac{\partial q_t}{\partial z}=\frac{\partial q_t}{\partial q}+i\frac{\partial q_t}{\partial p} \nonumber \\
Y=2\frac{\partial p_t}{\partial z}=\frac{\partial p_t}{\partial q}+i\frac{\partial p_t}{\partial p} \ .
\end{eqnarray}

The variational system has a unique global solution, bounded away from zero for finite time $t$ \cite{naz2}. The initial values are (suppressing the phase space dependence), $X(0)=I$ and $Y(0)=iI$. With respect to the variational system, the anisotropy form $Z$ is given by 
\begin{equation}
Z(t)=Y(t)X(t)^{-1} \ .
\end{equation}
The initial condition $G_{(q,p)}^Z(x,0;\hbar)=G_{(q,p)}(x;\hbar)$ determines the initial value $Z(0)=iI$, which is compatible with $Z(0)=Y(0)X(0)^{-1}=iI$, so that there is an arbitrariness in the initial values of matrices $X$ and $Y$ separately.

The dynamical representation of $Z$ is, then,
\begin{equation}
\hspace{-20mm}Z=YX^{-1}=\Big(\partial_qp_t+i\partial_pp_t\Big)\Big(\partial_qq_t+i\partial_pq_t\Big)^{-1}=\frac{\partial p_t}{\partial z}\left(\frac{\partial q_t}{\partial z}\right)^{-1} \ .
\label{eq:Zdyn}
\end{equation}

For Hamiltonians of the standard form, $H(q,p)=p^2+V(q)$, the variational system becomes
\begin{equation}
\frac{d}{dt}\left(
\begin{array}{ccc}
X \\
Y 
\end{array}\right)=\left(
\begin{array}{ccc}
0 &  -V''(q_t) \\
2I & 0  
\end{array}\right)\left(
\begin{array}{ccc}
X \\
Y 
\end{array}\right) \ ,
\end{equation}
and the matrix Riccati equation
\begin{equation}
\dot Z+2Z^2=-V''(q_t) \ .
\end{equation} 
For bounded motion (compact energy shell) and smooth potential, we have the following large time asymptotics,
\begin{eqnarray}
\dot Z_2\rightarrow  0 \ , \ \ Z_1Z_2+Z_2Z_1\rightarrow 0 \ , \nonumber \\
\ \dot Zq_t\sim\frac{1}{2}V'(q_t) \ ,
\end{eqnarray}
in matrix norm. Finally, the amplitude becomes
\begin{eqnarray}
\hspace{-20mm}u(q,p,t;\hbar)=(\pi\hbar)^{-d/4}\Big({\rm det}\,Z_2(q,p,t)\Big)^{1/4}e^{i\phi(q,p,t)}\nonumber \\
=(\pi\hbar)^{-d/4}\exp\Big(-\int_0^t{\rm tr} \, Z(q,p,\tau)\,d\tau\Big) \ .
\end{eqnarray}

\subsection{The Metaplectic Structure of the Anisotropic Gaussian Approximation}

The dynamics of the anisotropy form as defined by the anisotropic Gaussian wave packet approximation, is underlied by a rich algebraic and geometric structure. The semiclassical evolution of $G^Z$ is a representation of a one-parameter flow of $Z$ in $\mathbb{H}_d$, \textit{the  Weil representation of the metaplectic group ${\rm Mp}(2d,\mathbb{R})$ acting on} $\mathbb{H}_d$, intimately related to the nearby orbit approximation, as described above \cite{litt1,stoyanovsky}.

As we described above, the nearby orbit method effectively approximates the Hamitlonian flow by the linear flow generated by the quadratic Hamiltonian
\begin{equation}
H_0(q,p,t)=\frac{1}{2}\left(\begin{array}{ccc}
q  \\
p   
\end{array}\right)^T   H''(g^t(q_0,p_0))\left(\begin{array}{ccc}
q  \\
p   
\end{array}\right)
 \ , 
\end{equation}
fixed on a reference point $g^t(q_0,p_0)$. Under the quantized linearized flow induced by the above quadratic Hamiltonian, one may construct an \textit{exact} solution of the Schr\"{o}dinger equation with initial data
\begin{equation}
\mathcal{G}^Z_{\kappa}(x;\hbar)=(\pi\hbar)^{-d/4}\Big(\det \,{\rm Im}\,Z\Big)^{1/4}e^{\frac{i}{\hbar}\Big(\kappa\cdot x+\frac{1}{2}x\cdot Zx\Big)} \ .
\end{equation}
If the linear flow, $g^t_0$, is given by
\begin{equation}
g_0^t=\left(
\begin{array}{ccc}
A& B \\
C & D \end{array}\right) \in {\rm Sp}(2d,\mathbb{R})\ ,
\end{equation}
then the exact solution is \cite{stoyanovsky}, 
\begin{equation}
\Big({\rm det}\,(CZ+D)\Big)^{-1/2}e^{-\frac{i}{2\hbar}\kappa\cdot (CZ+D)^{-1}C\kappa}\,\mathcal{G}^{f(Z)}_{[(CZ+D)^T]^{-1}\kappa} \ ,
\end{equation}

where $f(Z)=(AZ+B)(CZ+D)^{-1}$, with $\det (CZ+D)\neq 0$. 

This defines the action of the symplectic group on the set of Gaussian wave packets, which simultaneously guarantees that the generalized M\"{o}bius automorphism on $\mathbb{H}_d$
\begin{equation}
Z\mapsto f(Z)=(AZ+B)(CZ+D)^{-1}  \ , \ \ \det\,(CZ+D)\neq 0 \ ,
\end{equation}
is a group action of $\sigma\in {\rm Sp}(2d,\mathbb{R})$ on $\mathbb{H}_d$.

Thus, the class of anisotropic Gaussian wave packets, 
\begin{equation} 
\mathcal{G}^Z_{\kappa}(x;\hbar)=(\pi\hbar)^{-d/4}\Big(\det \,{\rm Im}\,Z\Big)^{1/4}e^{\frac{i}{\hbar}\Big(\kappa\cdot x+\frac{1}{2}x\cdot Zx\Big)} \ , 
\end{equation}
which are obviously square integrable, 
\begin{equation}
\mathcal{G}^Z_\kappa\in L^2(\mathbb{R}^d) \iff  Z_@\succ 0 \ \ {\rm and} \ \ \kappa\in\mathbb{R}^d \ ,
\end{equation}
transform into one another under the action of the symplectic group \cite{stoyanovsky}
\begin{equation}
\hspace{-20mm}\mathcal{G}^Z_\kappa\mapsto M(\sigma)\mathcal{G}^Z_\kappa= \Big({\rm det}\,(CZ+D)\Big)^{-1/2}e^{-\frac{i}{2\hbar}\kappa\cdot (CZ+D)^{-1}C\kappa}\,\mathcal{G}^{f(Z)}_{[(CZ+D)^T]^{-1}\kappa} \ .
\label{eq:spwp}
\end{equation}

Although closely related to the symplectic group, the group action ($\ref{eq:spwp}$) is not the representation of ${\rm Sp}(2d,\mathbb{R})$ on $L^2(\mathbb{R}^d)$ Gaussian wave packets, due to the double valuedness induced by the square root of the determinant factor \cite{litt1,stoyanovsky}. 

In order to achieve a single valued representation, one must consider \textit{the  metaplectic group}, which is the double covering group of the symplectic group, taking into account exactly this ambiguity. By representing a member of the symplectic group $\sigma=\left(\begin{array}{ccc}
A & B \\
C & D  
\end{array}\right)\in {\rm Sp}(2d,\mathbb{R})$ for appropriate $A,B,C$ and $D\in\mathbb{C}^{d\times d}$ so that $\sigma J\sigma^T=J$, we define the metaplectic group as
\begin{equation}
{\rm Mp}(2d,\mathbb{R})=\Big\{(\sigma,\iota):\sigma \in{\rm Sp}(2d,\mathbb{R}) \ , \ \ \iota=\pm1\Big\} \ ,
\end{equation}
the choice of the index $\iota$ being a choice of either branch of the square root of the determinant $\Big(\det\,(CZ+D)\Big)^{1/2}$ in the complex plane, for $Z\in\mathbb{H}_d$. \begin{equation} 
(\sigma_1,\iota_1)\cdot (\sigma_2,\iota_2):=M_{\iota_1}(\sigma_1)M_{\iota_2}(\sigma_2)=M_{\iota_1\iota_2}(\sigma_1\sigma_2) \ .
\end{equation}

The representation of ${\rm Mp}(2d,\mathbb{R})$ on $L^2(\mathbb{R}^d)$ is called \textit{the  Weil representation}. Its action on Gaussian wave packets acquires the above simple form; in general, given that the block $B$, say, is nonsingular, the corresponding general metaplectic action on $L^2(\mathbb{R}^d)$ is given by \cite{litt1}
\begin{equation}
\hspace{-24mm}M_{\pm}(\sigma)\psi(x)=\int\frac{\pm1}{(2\pi i\hbar)^{d/2}\sqrt{|\det\, B}|}e^{\frac{i}{2\hbar}\Big(y\cdot B^{-1}Ay-2y\cdot B^{-1}x+x\cdot DB^{-1}x\Big)}\psi^\hbar(y)\,dy \ ,
\end{equation}
assuming, without loss of generality that $\det\, B\neq 0$.

The above uncovers the following interconnection: semiclassical anisotropic Gaussian wave packet dynamics is equivalent to a particular matrix Riccati dynamics of the anisotropy quadratic form, which is identified with a one-parameter flow of ${\rm Mp}(2d,\mathbb{R})$ in $\mathbb{H}_d$, which in turn, is equivalent to the stability dynamics of the Hamiltonian flow.

By establishing an approximate solution for the propagation of wave packets, we can use the linearity of the Schr\"{o}dinger propagation and the wave packet transform to construct approximate solutions of the Schr\"{o}dinger equation, and in turn, of the phase space Schr\"{o}dinger equation, with any initial data. The idea is trivial, intuitevely; approximate the evolution of a general initial wave function by evolving the wave packets into which it is analyzed, and superpose them back together to obtain the asymptotic solution.

\subsection{Anisotropic Wave Packet Dynamics in Phase Space}

Starting from ($\ref{eq:KG}$), we note that a basic step toward the construction of a semiclassical wave packet propagator based on the anisotropic wave packet dynamics, 
\begin{equation}
U^tG_{(q,p)}(x;\hbar)\sim G^Z_{(q,p)}(x,t;\hbar) \ ,
\end{equation}
is the transformation of wave packets into the phase space under $\mathcal{W}$. In particular, one must guarantee that the anisotropic wave packet dynamics retains, under $\mathcal{W}$, its Gaussian form.

The wave packet transform of the anisotropic Gaussian wave packet reads 
\begin{eqnarray}
\hspace{-25mm}\mathcal{W}G^Z_{(\eta,\xi)}(q,p,t;\hbar)=\Big(\frac{1}{2\pi\hbar}\Big)^{d/2}\int \bar G_{(q,p)}(x;\hbar)G^Z_{(\eta,\xi)}(x,t;\hbar)\,dx \nonumber \\
\hspace{-25mm}=(\pi\hbar)^{-d/2}\Big({\rm det}\,Z_2(\eta,\xi,t)\Big)^{1/4} \sqrt{\Big|\det\, W(\eta,\xi,t)\Big|}e^{i\lambda(\eta,\xi,t)} \nonumber \\
\hspace{-25mm}\times \exp\frac{i}{\hbar}\Big(A(\eta,\xi,t)-p\cdot(\eta_t-q)+\frac{i}{2}(\eta_t-q)^2-\frac{1}{2}(z-\zeta_t)\cdot  W(\eta,\xi,t)(z-\zeta_t)\Big) \ ,
\label{eq:wgz}
\end{eqnarray}
where $z=q-ip$, $\zeta=\eta-i\xi$, and
\begin{equation}
W(\eta,\xi,t):=i\partial_\zeta\eta_t(\partial_\zeta\zeta_t)^{-1}=-\Big(Z(\eta,\xi,t)+iI\Big)^{-1} \ .
\label{eq:w}
\end{equation}
The overall phase is 
\begin{eqnarray}
\lambda(\eta,\xi,t):=\phi(\eta,\xi,t)+\frac{1}{2}{\rm Arg}\,{\rm det}\,W(\eta,\xi,t)-\frac{\pi d}{4}\nonumber\\
=-\frac{1}{2}\int_0^t{\rm tr}\,H_{\xi\xi}''Z_2(\eta,\xi,\tau)\, d\tau +\frac{1}{2}{\rm Arg}\,{\rm det}\,W(\eta,\xi,t) -\frac{\pi d}{4}\ , 
\end{eqnarray}
by ($\ref{eq:u}$). The choice of the principal branch of the square root of the term $\Big({\rm det}\, W\Big)^{1/2}$ is necessary in order to ensure agreement with the initial data $\mathcal{W}G^Z_{(\eta,\xi)}(q,p,0;\hbar)=\mathcal{W}G_{(\eta,\xi)}(q,p;\hbar)$, as $\lambda(\eta,\xi,0)=0$, which in turn ensures agreement with generic initial data in for the action of the semiclassical quantum flow. 

This choice of phase actually has a nontrivial algebraic content; the double valuedness of the square root of the determinant reflects the double valuedness of the covering mapping ${\rm Sp}(2d,\mathbb{R})\rightarrow {\rm Mp}(2d,\mathbb{R})$, and thus choice of branch is actualy a choice of a specific metaplectic transformation.

In order to verify the invariance of Gaussian form under $\mathcal{W}$, as noted by Folland \cite{folland}, i.e., that wave packets in configuration space are mapped into wave packets in phase space, one must bring the above in the form of a Gaussian with a Siegel quadratic form in $(q,p)$, rather than in $z=q-ip$.

The auxiliary mapping $Z\mapsto W= f (Z)=-(Z+iI)^{-1}$, which will serve in defining the phase space anisotropy form as will be shown, is an action of the metaplectic group on $\mathbb{H}_d$, and its image is a submanifold of the open Siegel upper half disk $\mathbb{D}_d^+$,
\begin{equation}
f (\mathbb{H}_d)\subset \mathbb{D}_d^+=\Big\{W\in\mathbb{H}_d:\, I-W^*W\succ 0\Big\} \ ,
\end{equation}
and is actually identified with the open ball in the matrix norm topology,
\begin{equation}
f(\mathbb{H}_d)=B_{1/2}\Big(\frac{i}{2}I\Big)=\Big\{W\in\mathbb{H}_d:\,  \|W-\frac{i}{2}I\|_{\mathbb{C}^{d\times d}}<\frac{1}{2} \Big\} \ .
\end{equation}
The closed ball is the image of the compactified Siegel half space, $\bar \mathbb{H}_d$, extending it by the real symmetric matrices (the boundary) and the point at infinity. Thus, we have the parametrization of the boundary by the matrix unitary group ${\rm U}(d)$, 
\begin{equation}
\partial f(\bar \mathbb{H}_d)=\Big\{W(U)=\frac{i}{2}I+U:\,U\in {\rm U}(d)\Big\} \ .
\end{equation}

In the light of this geometric restriction, the above is brought in a more transparent form by expressing the phase of $\mathcal{W}G^Z$ in $(\ref{eq:wgz})$ as a \textit{real} quadratic form over phase space,
\begin{eqnarray}
\hspace{-20mm}\mathcal{W}G^Z_{(\eta,\xi)}(q,p;\hbar)=(\pi\hbar)^{-d/2}\Big(\det  Q _2(\eta,\xi,t)\Big)^{1/4}e^{i\lambda(\eta,\xi,t)}\nonumber \\
\hspace{-20mm}\times \exp\frac{i}{\hbar}\Big(A(\eta,\xi,t)-p\cdot (\eta_t-q)+\frac{1}{2}\left(\begin{array}{ccc}
q-\eta_t\\
p-\xi_t  
\end{array}\right)^T  
 Q(\eta,\xi,t)  \left(
\begin{array}{ccc}
q-\eta_t\\
p-\xi_t  
\end{array}\right)\Big) \ .
\end{eqnarray}
The matrix of the quadratic form is 
\begin{equation}
Q =\left(\begin{array}{ccc}
iI-W & iW \\
iW & W  
\end{array}\right) \in\mathbb{C}^{2d\times 2d}\ , 
\end{equation}
the matrix $W$ as defined in $(\ref{eq:w})$.

This makes explicit the phase space Gaussian wave packet character of $\mathcal{W}G^Z_{(\eta,\xi)}(q,p)$, under the condition that the phase space anisotropy form guarantees square integrability, i.e., $Q_2\succ 0$. By noting, additionally, by the Plancherel relation $\|\mathcal{W}G^Z_{(\eta,\xi)}\|_{\mathfrak{F}^2}=\|G^Z_{(\eta,\xi)}\|_{L^2(\mathbb{R}^d)}=1$, that 
\begin{equation}
\Big({\rm det}\,Z_2\Big)^{1/4}\sqrt{\Big|{\rm det}\, W\Big|}=\Big({\rm det}\, Q_2\Big)^{1/4} \ ,
\end{equation} 
where $Q_2={\rm Im}\, Q$, the Gaussian form of $\mathcal{W}G^Z$ becomes more apparent.

Consider the mapping of the anisotropy form in phase space, 
\begin{equation}
h:\mathbb{H}_d\rightarrow \mathbb{C}^{2d\times 2d}|\,W\mapsto Q \ ,
\end{equation} 
as above. Since $h$ is continuous, and since $Q^T=Q$, all that is required in order to show that ${\rm im}\,h=\mathbb{H}_{2d}$ is to show that there are no points on the boundary $Q(U)\in\partial h\Big(f(\bar \mathbb{H}_d)\Big)=h\Big(\partial f(\bar \mathbb{H}_d)\Big)$, such that $Q_2(U)=0$, where the manifold $\partial h\Big(f(\bar \mathbb{H}_d)\Big)$ is parameterized by the equation
\begin{equation}
Q=h\Big(\frac{i}{2}I+U\Big) \ , \ \ U\in {\rm U}(d) \ .
\end{equation}
This is indeed true, as the matrix
\begin{equation}
Q_2(U)=\left(\begin{array}{ccc}
\frac{1}{2}I-U_2 & U_1 \\
U_1 & \frac{1}{2}I+U_2  
\end{array}\right) \ ,
\end{equation}
where $U=U_1+iU_2\in{\rm U}(d)$, is always nonzero. Thus, we see that the image of $f(\mathbb{H}_d)$ is mapped into a true subdomain of $\mathbb{H}_{2d}$ under $h$, and so the dynamics of $Q$ retains its Siegel character. Thus, $\mathcal{W}G^Z_{(\eta,\xi)}(q,p,t;\hbar)$ is a true phase space wave packet with anisotropy form $Q$.

This makes transparent the following geometric picture: \textit{the  representation of the metaplectic group ${\rm Mp}(2d,\mathbb{R})$ into $L^2(\mathbb{R}^d)$ is in 1-1 correspondence with the representation of the metaplectic group ${\rm Mp}(4d,\mathbb{R})$ into Fock-Bargmann space, $\mathfrak{F}^2$, by the wave packet transform}. In other words, the wave packet transformation of wave packets are true phase space wave packets themselves.

The fact that a Gaussian wave packet retains its form in phase space is reinforced by the fact that its Wigner transform, is in turn a real Gaussian wave packet, bearing in mind that the modulus square of the phase space wave function is the convolution of the Wigner function (see section (\ref{wignr}) of the appendix) with a localized Gaussian \cite{litt1}
\begin{equation}
\hspace{-20mm}W_{G_{(\eta,\xi)}}(q,p;\hbar)=(\pi\hbar)^{-d/2}\exp-\frac{1}{\hbar}\left(\begin{array}{ccc}
q-\eta_t\\
p-\xi_t  
\end{array}\right)^T  
 R(\eta,\xi,t)  \left(
\begin{array}{ccc}
q-\eta_t\\
p-\xi_t  
\end{array}\right) \ ,
\end{equation}
where $ R\in\mathbb{R}^{2d\times 2d}$ defines a symmetric and positive definite quadratic form, related to $Q$.

As in the case of the anisotropic Gaussian wave packet in configuration space, $G^Z$, the amplitude of its phase space image, $\mathcal{W}G^Z$, is controlled essentially by the quadratic term, which vanishes on the set $\Big\{(q,p,\eta,\xi):\, (q,p)=g^t(\eta,\xi)\Big\}$, so that it becomes semiclassically concentrated there, exponentially small at a fixed distance away.

\section{Semiclassical Propagator for the Phase Space Schr\"{o}dinger Equation}

\subsection{Construction of the Semiclassical Propagator}

We now construct a semiclassical propagator for the phase space Schr\"{o}dinger equation, \textit{the semiclassical wave packet propagator}, and a semiclassical flow, $\mathcal{U}^t_{wp}=\mathcal{W}\circ U_{wp}^t\circ \mathcal{W}^{-1}$, which is based on the anisotropic Gaussian wave packet approximation. 

By the inverse wave packet transform ($\ref{eq:iwpt})$ we have 
\begin{eqnarray}
\hspace{-20mm}\psi_0^\hbar (x)=\mathcal{W}^{-1}\Psi_0^\hbar (x)= \Big(\frac{1}{2\pi\hbar}\Big)^{d/2}\int G_{(q,p)}(x;\hbar)\Psi^\hbar_0(q,p)\,dqdp \ ,
\end{eqnarray}
and combining with ($\ref{eq:Kprop})$,  we get 
\begin{eqnarray}
\hspace{-20mm}\psi^\hbar(x,t)=\Big(\frac{1}{2\pi\hbar}\Big)^{d/2}\int \!\!\!\int K(x,y,t;\hbar)G_{(q,p)}(y;\hbar)\Psi^\hbar_0(q,p)\,dydqdp \ .
\end{eqnarray}
As $ G^Z _{(q,p)}$ is a semiclassical solution of the initial value problem  ($\ref{eq:ivp})$ with initial data $ \psi_0^\hbar (x)=G_{(q,p)}(x;\hbar)$, we have the approximation
\begin{equation}
\hspace{-20mm}U^t G_{(q,p)}(x;\hbar)=\int K(x,y,t;\hbar)G_{(q,p)}(y;\hbar)\,dy \sim G^Z _{(q,p)}(x,t;\hbar) \ .
\end{equation}
This, which is essentially the thawed Gaussian approximation, is the central approximation we make to construct the semiclassical phase space propagator. 

The last two equations yield the approximate solution 
\begin{equation}
\hspace{-25mm}\psi^\hbar(x,t)\sim \Big(\frac{1}{2\pi\hbar}\Big)^{d/2}\int G^Z _{(q,p)}(x,t;\hbar)\Psi^\hbar_0(q,p)\,dqdp =: \int K_{wp}(x,y,t;\hbar)\psi_0^\hbar(y)\,dy \ .
\end{equation}
The kernel $K_{wp}$ is expressed as 
\begin{equation}
K_{wp}(x,y,t;\hbar):=\Big(\frac{1}{2\pi\hbar}\Big)^{d}\int\bar G_{(q,p)}(y;\hbar) G^Z _{(q,p)}(x,t;\hbar)\,dqdp \ ,
\end{equation}
the kernel of the semiclassical flow $U^t_{wp}$.

We are interested in a semiclassical approximation for the phase space propagator. By substituting the approximate propagator $K_{wp}$ into the representation formulae ($\ref{eq:pswf}$) and ($\ref{eq:psprop}$), we obtain 
\begin{eqnarray}
\hspace{-20mm}\mathcal{U}^t\Psi^\hbar_0(q,p)\sim\mathcal{U}^t_{wp}\Psi^\hbar_0(q,p)
=:\int \mathcal{K}_{wp}(q,p,\eta,\xi,t;\hbar)\Psi^\hbar_0(\eta,\xi)\,d\eta d\xi\nonumber \\
=\Big(\frac{1}{2\pi\hbar}\Big)^{d}\int\!\!\!\int\bar G_{(q,p)}(x;\hbar) G^Z _{(\eta,\xi)}(x,t;\hbar)\Psi^\hbar_0(\eta,\xi)\,dxd\eta d\xi \ .
\end{eqnarray}
We term this \textit{the semiclassical wave packet approximation} for the propagator of the phase space Schr\"{o}dinger equation. 

The kernel of the semiclassical quantum flow $\mathcal{U}^t_{wp}$ is 
\begin{eqnarray}
\hspace{-20mm}\mathcal{K}_{wp}(q,p,\eta,\xi,t;\hbar):=\Big(\frac{1}{2\pi\hbar}\Big)^{d}\int\bar G_{(q,p)}(x;\hbar) G^Z _{(\eta,\xi)}(x,t;\hbar)\,dx \ ,
\end{eqnarray}
and is proportional to $\mathcal{W}G^Z$,
\begin{equation}
\mathcal{K}_{wp}(q,p,\eta,\xi,t;\hbar)=\Big(\frac{1}{2\pi\hbar}\Big)^{d/2}\mathcal{W}G^Z_{(\eta,\xi)}(q,p,t;\hbar) \ ,
\end{equation}
so that from $(\ref{eq:wgz})$ we finally deduce
\begin{eqnarray}
\hspace{-20mm}\mathcal{K}_{wp}(q,p,\eta,\xi,t;\hbar)=2^{-d/2}(\pi\hbar)^{-d}\Big({\rm det}\, Q _2(\eta,\xi,t)\Big)^{1/4} e^{i\lambda(\eta,\xi,t)}\nonumber \\
\hspace{-20mm}\times \exp\frac{i}{\hbar}\Big(A(\eta,\xi,t)-p\cdot (\eta_t-q)+\frac{1}{2}\left(\begin{array}{ccc}
q-\eta_t\\
p-\xi_t  
\end{array}\right)^T  
 Q(\eta,\xi,t)  \left(
\begin{array}{ccc}
q-\eta_t\\
p-\xi_t  
\end{array}\right)\Big) \ .
\label{eq:Lwp}
\end{eqnarray}

The semiclassical propagator $\mathcal{K}_{wp}$ is of the form of an approximate transition amplitude between a quantum state microlocalized at the phase space point $(q,p)$ and another moving with a time decay along the orbit $\gamma^t(\eta,\xi)$, semiclassically localized on the manifold $\Big\{(q,p,\eta,\xi):\, (q,p)=g^t(\eta,\xi)\Big\}$.

By construction, the semiclassical propagator is a flow, in the group sense, over the set of wave packets (in configuration space),
\begin{equation}
\hspace{-20mm}U^t_{wp}\circ U^{ t'}_{wp}G_{(q,p)}(x;\hbar)=U^t_{wp}G_{(q,p)}^Z(x, t';\hbar)=G^Z_{(q,p)}(x,t+ t';\hbar) \ ,
\end{equation}
for $t, t'\in\mathbb{R}$, which can be generalized to an arbitrary state by the overcompleteness of coherent states. 

Thus, we have that the approximate dynamics in phase space enduced by the operator $\mathcal{U}^t_{wp}$, besides satisfying the Fock-Bargmann constraints, is a unitary quantum flow in $\mathfrak{F}^2$,
\begin{equation}
\mathcal{U}_{wp}^{t \, *}=\mathcal{U}_{wp}^{-t} \ , \ \ \mathcal{U}_{wp}^t\circ\mathcal{U}^{t'}=\mathcal{U}_{wp}^{t+t'} \ , \ \ \mathcal{U}^0_{wp}={\rm Id}_{\mathfrak{F}^2} \ , \ \ t,t'\in\mathbb{R} \ .
\end{equation}
In terms of the semiclassical wave packet phase space propagator, we have
\begin{equation}
\hspace{-20mm}\int\mathcal{K}_{wp}(q,p,u,v,t;\hbar)\mathcal{K}_{wp}(u,v,\eta,\xi, t';\hbar)\,dudv=\mathcal{K}_{wp}(q,p,\eta,\xi,t+ t';\hbar) \ ,
\end{equation}
and 
\begin{equation}
\bar\mathcal{K}_{wp}(\eta,\xi,q,p,-t;\hbar)=\mathcal{K}_{wp}(q,p,\eta,\xi,t;\hbar) \ .
\end{equation}

At $t=0$, the semiclassical wave packet phase space propagator shares the reproducing property of the exact propagator,
\begin{eqnarray} 
\hspace{-20mm}\lim_{t\rightarrow 0^+}\mathcal{K}_{wp} (q,p,\eta,\xi,t;\hbar)=\Big(\frac{1}{2\pi\hbar}\Big)^{d}\int\bar G_{(q,p)}(x;\hbar) G^Z _{(\eta,\xi)}(x,0;\hbar)\,dx\nonumber \\
=\Big(\frac{1}{2\pi\hbar}\Big)^{d}\int\bar G_{(q,p)}(x;\hbar) G_{(\eta,\xi)}(x;\hbar)\,dx=\mathcal{K}(q,p,\eta,\xi,0;\hbar) \ .
\end{eqnarray}

\subsection{The Nearby Orbit Approximation and Initial Value Representations}

In \cite{litt1}, Littlejohn constructed a semiclassical propagator based on the nearby orbit approximation, as an explicit action of Weyl and metaplectic operators, generalizing the nearby orbit approximation for the dynamics of Liouville densities, in the quantum mechanical framework. 

This construction, however, does not fall in the category of initial value representations. The Littlejohn propagator utilizes a single reference orbit, with starting point the phase space point on which the initial data $\psi_0^\hbar$ is assumed to be centered at and localized, without necessarily being Gaussian. Although reminescent of single semiclassical Gaussian wave packet propagation, it loses unitarity and linearity as the base point $(q ,p )$ depends implicitly on $\psi_0^\hbar$. 

The semiclassical Littlejohn flow in the nearby orbit approximation reads 
\begin{equation}
U_{nbo}^t\psi^\hbar(x):=e^{\frac{i}{\hbar}\theta(z ,t)}\mathcal{T}_{z_t}\circ M(\sigma_t)\circ\mathcal{T}^*_{z }\psi^\hbar(x) \ ,
\end{equation}
where $\mathcal{T}_z:=e^{\frac{i}{\hbar}\Big(p\cdot \hat x-q\cdot \hat p_x\Big)}$ is the Weyl operator (see subsection (\ref{twpt}) of the appendix), and $z_t=q_t-ip_t$ defines the reference orbit with initial point $z =q -ip $. The ambiguity of sign of the metaplectic operator is resolved by the requirement that $U_{nbo}^0={\rm Id}_{L^2(\mathbb{R}^d)}$, so that $M(\sigma):=M_+(\sigma)$, while the symplectic transformation $\sigma_t=\left(
\begin{array}{ccc}
A_t &  B_t \\
C_t & D_t  
\end{array}\right)\in{\rm Sp}(d,\mathbb{R})$, is the solution of the initial value problem 
\begin{eqnarray}
\frac{d\sigma_t}{dt}=J''H(g^t(q,p))\sigma_t \nonumber \\
\sigma_0=I \ ,
\end{eqnarray}
and the additional phase is $\theta(z ,t)=A(q ,p ,t)-\frac{1}{2}(q_t\cdot p_t-q \cdot p )$.

The propagator fixes the initial state from $(q ,p )$ to the origin, symplectically `rotates' it in phase space by the action of $M(\sigma_t)$, and shifts it along the reference orbit modulating by adding the action phase, modulating with the extra action phase. Explicitly, we have 
\begin{eqnarray}
\hspace{-20mm}U^t_{nbo}\psi^\hbar_0(x)=\Big(\frac{1}{2\pi\hbar}\Big)^{d/2}\frac{e^{\frac{i}{\hbar}\Big(A(q ,p ,t)+\frac{1}{2}(y-q )\cdot B^{-1}_tA_t(y-q )\Big)}}{\sqrt{|\det\, B_t|}}\nonumber  \\
\hspace{-20mm}\times\int\exp\frac{i}{\hbar}\Big(-(y-q )\cdot B^{-1}_t(x-q_t)+\frac{1}{2}(x-q_t)\cdot D_tB^{-1}_t(x-q_t)\Big)\psi_0^\hbar(y)\, dy \ .
\end{eqnarray}

Although the method of constructing the semiclassical propagator $\mathcal{K}_{wp}$ in the previous subsection relies on the nearby orbit scheme as well, there is a number of differences to the Littlejohn semiclassical propagator. 

The Littlejohn semiclassical propagator is simpler computationally, as it involves a single orbit eminating from the center of a microlocalized initial state. Subsequently, it does not grasp the quantum dynamics for nonmicrolocalized initial data; in the anisotropic wave packet approximation elaborated on previously, one can consider initial Gaussians for all possible initial phase space points, and thus all possible reference orbits in phase space. This renders the use of semiclassical wave packet propagator valid even in the case of nonmicrolocalized initial data, such as WKBM states, concentrated on a certain Lagrangian manifold. The other differences have to do with the linearity and unitarity properties.

A different approach, on the lines of which we constructed the semiclassical wave packet propagator, is that of initial value representations for solutions of the Schr\"{o}dinger equation. By beginning with the completeness relation of the isotropic Gaussian states in phase space $(\ref{eq:completeness})$,
\begin{equation}
\hspace{-20mm}K(x,y,t;\hbar)=\Big(\frac{1}{2\pi\hbar}\Big)^{d} \int U^tG_{(q,p)}(x;\hbar)\bar G_{(q,p)}(y;\hbar)\,dqdp \ ,
\end{equation}
where $K$ is the Schr\"{o}dinger propagator, the following approximate phase space resolutions are usually made in this context,
\begin{eqnarray}
\hspace{-20mm}K(x,y,t;\hbar)\sim \Big(\frac{1}{2\pi\hbar}\Big)^{d} \int G^Z_{(q,p)}(x;\hbar)\bar G_{(q,p)}(y;\hbar)\,dqdp\nonumber \\
\hspace{-20mm}K(x,y,t;\hbar)\sim \Big(\frac{1}{2\pi\hbar}\Big)^{d} \int e^{\frac{i}{\hbar}S(q,p,t)}G_{(q,p)}(x;\hbar)\bar G_{(q,p)}(y;\hbar)\,dqdp\nonumber \\
\hspace{-20mm}K(x,y,t;\hbar)\sim \Big(\frac{1}{2\pi\hbar}\Big)^{d} \int e^{\frac{i}{\hbar}S(q,p,t)}a(q,p,t;\hbar)G_{(q,p)}(x;\hbar)\bar G_{(q,p)}(y;\hbar)\, dqdp \ .
\end{eqnarray}
The above, are respectively \textit{the thawed Gaussian approximation}, \textit{the frozen Gaussian approximation}, and \textit{the Herman-Kluk approximation} \cite{HK,rober HK, rousse}.

\subsection{Relation to the van Vleck-Gutzwiller Propagator}

We derive asymptotics for the Schr\"{o}dinger representation of the semiclassical wave packet propagator by means of the complex stationary phase method (see subsection $(\ref{stplemma})$ of the appendix). 

As was previously noted, 
\begin{equation}
K_{wp}(x,y,t;\hbar):=\Big(\frac{1}{2\pi\hbar}\Big)^{d} \int\bar G_{(q,p)}(y;\hbar) G^Z _{(q,p)}(x,t;\hbar)\,dqdp \ .
\end{equation}
Bringing the above in the standard oscillating integral form, we have 
\begin{equation}
K_{wp}(x,y,t;\hbar)=\int \chi(q,p,t;\hbar)e^{\frac{i}{\hbar} \Phi(q,p,x,y,t)}\,dqdp \ ,
\end{equation}
where the phase and amplitude read 
\begin{eqnarray}
\hspace{-20mm}\Phi(q,p,x,y,t)=A(q,p,t)+p\cdot (q-y)+\frac{i}{2}(q-y)^2\nonumber \\
+p_t\cdot (x-q_t)+\frac{1}{2}(x-q_t)\cdot Z(x-q_t) \ ,
\end{eqnarray}
and 
\begin{equation}
\chi(q,p,t;\hbar)=2^{-d}(\pi\hbar)^{-3d/2} \Big({\rm det}\,Z_2(q,p,t)\Big)^{1/4} e^{i\phi}\ .
\end{equation}

The leading order contribution comes from the real critical manifold, 
\begin{equation}
\hspace{-20mm}\mathcal{C}_\Phi^\mathbb{R}=\Big\{(q,p):\,{\rm Im}\,\Phi(q,p,x,y,t)=0 \ , \ \ \nabla_{(q,p)}\Phi(q,p,x,y,t)=0\Big\} \ ,
\end{equation}
which is given by the equations 
\begin{equation}
\frac{\partial \Phi}{\partial q}(q,p,x,y,t)=0 \ , \ \ \frac{\partial \Phi}{\partial p}(q,p,x,y,t)=0 \ ,
\end{equation}
and
\begin{equation}
\frac{1}{2}(q-y)^2+\frac{1}{2}(x-q_t)\cdot Z_2 (x-q_t)=0 \ .
\end{equation}
As the matrix $Z_2$ is positive definite, the later is possible only if $y=q$ and $x=q_t$. Subsequently, the former stationary equations are, implying appropriate matrix muliplication,
\begin{equation}
p+\partial _qA-p_t\cdot \partial _qq_t=0 \ \ {\rm and} \ \ \partial _pA-p_t\cdot \partial _pq_t=0 \ ,
\end{equation} 
which by the properties of the phase space action $(\ref{eq:dA})$ are automatically satisfied.

Thus, the real critical manifold becomes 
\begin{equation}
\mathcal{C}^\mathbb{R}_{\Phi}=\Big\{(q,p):\, (q_t,q)=(x,y)\Big\} \ ,
\end{equation}
which corresponds to orbits, which, projected onto configuration space, yield rays beginning at $y$ and terminating at $x$. As there are more than one such possibilities, the critical set comprises of a countable set of rays.

On $\mathcal{C}^\mathbb{R}_{\Phi}$ the phase itself reduces to the action, while the restriction of its Hessian therein reads 
\begin{equation}
\hspace{-20mm}\Phi''_{(q,p)}=\left(
\begin{array}{ccc}
-\partial_qq_t\partial _qp_t+\partial_qq_t Z\partial _qq_t+iI &  -\partial _qp_t\partial _pq_t+\partial_q q_t Z\partial_pq_t \\
-\partial _qp_t \partial _pq_t+\partial_q q_t Z\partial_pq_t & -\partial_pq_t\partial _pp_t+\partial_pq_t Z\partial _pq_t  
\end{array}\right) \ ,
\end{equation}
by virtue of the matrix Poisson relation,
\begin{equation}
\frac{\partial q_t}{\partial q}\frac{\partial p_t}{\partial p}-\frac{\partial p_t}{\partial q} \frac{\partial q_t}{\partial p}=I \ .
\end{equation}

The determinant of the Hessian on the real critical manifold $\mathcal{C}_{\Phi}^\mathbb{R}$ gives 
\begin{equation}
{\rm det}_{\mathbb{C}^{2d\times 2d}} \Phi''_{(q,p)} = {\rm det}_{\mathbb{C}^{d\times d}}\bigg( \frac{1}{i}\partial _pq_t\cdot (\partial_zq_t)^{-1}\bigg) \ .
\end{equation}

Assuming short times, by the stationary phase lemma (in the weak sense),
\begin{equation}
\hspace{-20mm}K_{wp}(x,y,t;\hbar)\sim \frac{1}{(2\pi i\hbar)^{d/2}}\sum_{\gamma\in\Gamma_{y,x}^t}a_\gamma\sqrt{|\det\, \partial_{q_t}p|}e^{\frac{i}{\hbar}A_\gamma-\frac{\pi i}{2}\nu_{\gamma}} \ ,
\end{equation}
where $\pi$ is the canonical projection down to configuration space, and 
\begin{equation}
\Gamma_{y,x}^t:=\Big\{\gamma\in C^{2}(\mathbb{R}_+;\mathbb{R}^{2d}):\pi \gamma(0)=y \ , \ \ \pi\gamma(t)=x\Big\} 
\end{equation}
is the set of orbits with conjugate points at $y$ and $x$ at times $0$ and $t$ respectively. The phase $\nu_\gamma$ is the corresponding Maslov index of orbit $\gamma$,
\begin{equation}
\nu_\gamma={\rm ind}(\partial_{q_t}p):=\sum_{\lambda\in\sigma(\partial_{q_t}p)}{\rm Arg}(\lambda) \ ,
\end{equation}
i.e., the index of the monodromy matrix, the excess of its positive over negative eigenvalues, while $A_\gamma$ is the action of the orbit $\gamma$. The factor
\begin{equation}
a_\gamma=2^{d/2}\sqrt{\det\,\partial _zq_t}e^{-\frac{1}{2}\int_0^t{\rm tr} \,H_{pp}'' Z(q,p,\tau)\,d\tau} \ ,
\end{equation}
stands as a correction to the semiclassical van Vleck-Gutzwiller approximation of the Schr\"{o}dinger propagator. Indeed, for $t\rightarrow 0^+$, we have 
\begin{equation}
a_\gamma=1+O_{(q,p)}(t) \ .
\end{equation}

Additionally, in order to ensure that $K_{wp}$ serves as an approximation to the Schr\"{o}dinger propagator, we must have that the amplitude $a$ is nonsingular, even on the caustics. 

The exponential factor $\exp\Big(-\frac{1}{2}\int_0^t{\rm tr} \, H_{pp}'' Z(q,p,\tau)\,d\tau\Big)$ is always bounded due to the global existence of solutions of the matrix Riccati equation \cite{naz2}. As the flow $g^t$ constitutes a canonical transformation for all times, the determinant of $\partial q_t/\partial z$ is nonsingular, and thus $a_\gamma$ remains bounded both away from zero and infinity for all times.

Thus, we have shown, that for small times, the semiclassical wave packet propagator gives indeed the correct semiclassical asymptotics,
\begin{equation}
K_{wp}(x,y,t;\hbar)\sim \frac{1}{(2\pi i\hbar)^{d/2}}\sum_{\gamma\in\Gamma_{y,x}^t}\sqrt{\Big|\det\, \frac{\partial p}{\partial q_t}\Big|}e^{\frac{i}{\hbar}A_\gamma-\frac{\pi i}{2}\nu_{\gamma}} \ ,
\end{equation}
the semiclassical van Vleck-Gutzwiller approximation of the propagator.

\section{Asymptotic Solutions of the Phase Space Schr\"{o}dinger Equation}
\label{solutions}

We now turn to the issue of central interest, the asymptotic solutions for the semiclassical initial value problem of the phase space Schr\"{o}dinger equation. We construct such asymptotic solutions by semiclassically evolving appropriate phase space states under the semiclassical wave packet flow, $\mathcal{U}^t_{wp}$.

As we consider asymptotic solutions in the semiclassical approximation, it is necessary to consider \textit{semiclassical initial data} in phase space. To this end we consider the phase space image, under $\mathcal{W}$, of the WKBM state
\begin{equation}
\psi^\hbar_0(x)=a(x)e^{\frac{i}{\hbar}S(x)} \ .
\end{equation}

The phase is real valued and smooth, $S\in C^\infty (\mathbb{R}^d,\mathbb{R})$, while the amplitude is real valued and rapidly decrasing, $a\in \mathcal{S}(\mathbb{R}^d,\mathbb{R})$, satisfying, additionally a normalization condition
\begin{equation}
\int a(x)^2\, dx=1 \ .
\end{equation}

The phase space WKBM state is, for any $N\in\mathbb{Z}_+$ \cite{naz1}
\begin{eqnarray}
\hspace{-20mm}\Psi_0^\hbar(q,p)=\mathcal{W}\psi^\hbar_0(q,p)=a(q,p;\hbar)e^{\frac{i}{\hbar}\Sigma(q,p)}\nonumber\\
=i^{-d/2}(\pi\hbar)^{-d/4}e^{\frac{i}{\hbar}\Sigma(q,p)}\Bigg(\sum_{k=0}^{N-1}\hbar^ka_k(q,p)+O(\hbar^N)\Bigg) \ ,
\label{eq:wkb0}
\end{eqnarray}
where 
\begin{equation}
\Sigma(q,p):=\tilde S(\underline x(q,p))-p\cdot \Big(\underline x(q,p)-q\Big)+\frac{i}{2}\Big(\underline x(q,p)-q\Big)^2 \ , 
\label{eq:S}
\end{equation} 
and 
\begin{equation}
a_0(q,p)=\frac{\tilde a(\underline x(q,p))}{\sqrt{\det\Biggl(-\Big(\tilde S''(\underline x(q,p))+iI\Big)\Biggr)}} \ .
\end{equation}

The tilde stands for \textit{the almost analytic extension}, and $\underline x(q,p)$ stands for \textit{the almost analytic solution} of the stationary equation $\partial_w \tilde S(w)-p+i(w-q)=0$, where $w=x+iy\in\mathbb{C}^d$ is the complexification of $x$, (see subsection of the appendix (\ref{almost})), which reads,
\begin{equation}
\hspace{-20mm}\underline x(q,p)=q+\Big(S''(q)+iI\Big)^{-1}\Big(p-S'(q)\Big)+O\Bigg(\Big(p-S'(q)\Big)^2\Bigg) \ ,
\end{equation}
in a phase space neighborhood of the manifold $\Lambda_S$, the Lagrangian manifold of the initial phase $S$,
\begin{equation}
\Lambda_S=\Big\{(q,p):\,p=S'(q)\Big\} \ .
\end{equation}
The imaginary part of the phase $\Sigma(q,p)$ is non-negative, $\Sigma_2\geqslant 0$ \cite{naz1}, becoming zero only for $(q,p)\in\Lambda_S$, the Lagrangian manifold of the initial phase $S$, so that $\underline x|_{\Lambda_S}(q,p)=q$.

To leading order, the asymptotics of the wave packet transform of the initial data are
\begin{equation}
\hspace{-20mm}\Psi_0^\hbar(q,p)\sim (\pi\hbar)^{-d/4}\frac{i^{-3d/2}\tilde a(\underline x(q,p))}{\sqrt{\det\,\Big(\tilde S''(\underline x(q,p))+iI\Big)}}e^{\frac{i}{\hbar}\Sigma(q,p)} \ ,
\end{equation}
assuming the principal branch of the square root. Along the Lagrangian manifold, $(q,p)\in\Lambda_S$, we have 
\begin{equation}
\Psi_0^\hbar(q,p)\sim (\pi\hbar)^{-d/4}\frac{i^{-3d/2}a(q)}{\sqrt{\det\,\Big(S''(q)+iI\Big)}}e^{\frac{i}{\hbar}S(q)} \ .
\end{equation}

Considering the above class of phase space states as initial data, we construct the corresponding asymptotic solutions, $\Psi_{wp}^\hbar$, by the action of the semiclassical wave packet propagator
\begin{equation}
\Psi^\hbar_{wp}(q,p,t)=\int\mathcal{K}_{wp}(q,p,\eta,\xi,t;\hbar)\Psi_0^\hbar(\eta,\xi)\, d\eta d\xi \ ,
\end{equation}
which is asymptotic to the corresponding solution $\Psi^\hbar(q,p,t)$, in the sense that 
\begin{equation}
\Big(i\hbar\partial_t-\check H\Big)(\Psi_{wp}^\hbar-\Psi^\hbar)\rightarrow 0 \ , 
\end{equation}
weakly, algebraically in $\hbar$.

We examine the asymptotics of the integral representation of the asymptotic solution, according to the asymptotic series for $\Psi_0^\hbar$ in $(\ref{eq:wkb0})$, which we write in normal form (see subsection (\ref{stplemma}) of the appendix),
\begin{equation}
\Psi^\hbar_{wp}(q,p,t)\sim\int\chi(\eta,\xi,t;\hbar)e^{\frac{i}{\hbar}\Phi(q,p,\eta,\xi,t)}\,d\eta d\xi\ ,
\label{eq:assolut}
\end{equation}
where 
\begin{equation}
\hspace{-25mm}\chi(\eta,\xi,t;\hbar)=2^{-d/2}(\pi\hbar)^{-5d/4}\Big(\det\,Q_2(\eta,\xi,t)\Big)^{1/4}\,\frac{i^{-3d/2}\tilde a(\underline x(\eta,\xi))\,e^{i\lambda(\eta,\xi,t)}}{\sqrt{{\rm det}\,\Big(\tilde S''(\underline x(\eta,\xi))+iI\Big)}} \ ,
\end{equation}
and
\begin{eqnarray}
\hspace{-20mm}\Phi(q,p,\eta,\xi,t)=\tilde S(\underline x(\eta,\xi))-\xi\cdot \Big(\underline x(\eta,\xi)-\eta\Big)+\frac{i}{2}\Big(\underline x(\eta,\xi)-\eta\Big)^2+A(\eta,\xi,t)\nonumber\\
-p\cdot(\eta_t-q)+\frac{1}{2}\left(\begin{array}{ccc}
q-\eta_t\\
p-\xi_t  
\end{array}\right)^T  
 Q (\eta,\xi,t) \left(
\begin{array}{ccc}
q-\eta_t\\
p-\xi_t  
\end{array}\right) \ .
\end{eqnarray}

The critical manifold of the phase $\Phi$ in complexified phase space, is
\begin{equation}
\mathcal{C}_\Phi=\Big\{(z,\zeta):\, \nabla_{(z,\zeta)}\tilde \Phi(q,p,z,\zeta,t)=0\Big\} \ ,
\end{equation}
where $z=\eta+iu, \, \zeta=\xi+iv\in\mathbb{C}^{2d}$ are complexifications of the phase space canonical coordinates; the manifold $\mathcal{C}_\Phi$ is parameterized by the equations 
\begin{equation}
\eta=\underline z(q,p,t)  \ , \ \ \xi=\underline \zeta(q,p,t) \ .
\end{equation}

The general expression of the asymptotic solution is given by the complex stationary phase lemma (see subsection (\ref{stplemma}) of the appendix),
\begin{equation}
\hspace{-25mm}\Psi^\hbar_{wp}(q,p,t)\sim \Big(\frac{2\pi\hbar}{i}\Big)^d\,\frac{\tilde \chi(q,p,\underline z(q,p,t),\underline \zeta(q,p,t),t;\hbar)e^{\frac{i}{\hbar}\tilde \Phi(q,p,\underline z(q,p,t),\underline \zeta(q,p,t),t)}}{\sqrt{\det\Biggl(-\tilde \Phi''_{(z,\zeta)}(q,p,\underline z(q,p,t),\underline \zeta(q,p,t),t)\Biggr)}}  \ .
\end{equation}

As a phase space sumbmanifold, the nodal set of the imaginary part of the phase, $\Phi_2$ (see subsection (\ref{almost}) of the appendix), $\mathcal{Z}_{\Phi_2}=\Big\{(q,p):\,\Phi_2=0\Big\}$ is identified with the Lagrangian manifold \textit{of the propagated phase},
\begin{equation}
\mathcal{Z}_{\Phi_2}=g^t\Lambda_S=:\Lambda^t_S \ .
\end{equation}
This is due to the fact that $\Sigma_2\geqslant 0$ \cite{naz1}, and the positive definitiveness of the imaginary part of the phase space anisotropy form, $Q_2\succ 0$. Thus, the real critical manifold becomes 
\begin{eqnarray}
\hspace{-20mm}\mathcal{C}_{\Phi}^\mathbb{R}=\Big\{(\eta,\xi):\, \Phi_2(q,p,\eta,\xi,t)=0 \ , \ \ \nabla_{(\eta,\xi)}\Phi(q,p,\eta,\xi,t)=0\Big\}\nonumber \\
=\Big\{(\eta,\xi):\,(\eta,\xi)=g^{-t}(q,p)\in\Lambda_{S}\Big\}  \ ,
\end{eqnarray}
parameterized by the equations $(\eta,\xi)=g^{-t}(q,p)$, or
\begin{equation}
\eta=\underline q(-t;q,S'(q)) \ , \ \ \xi=\underline p(-t;q,S'(q)) \ .
\end{equation}

The asymptotic solution on the transported Lagrangian manifold $\Lambda^t_S$ becomes 
\begin{eqnarray}
\hspace{-25mm}\Psi_{wp}^\hbar(q,p,t)\sim(\pi\hbar)^{-d/4}\frac{i^{-5d/2}a(q_{-t})e^{i\lambda(q_{-t},p_{-t},t)}e^{\frac{i}{\hbar}\Phi(q,p,q_{-t},p_{-t},t)}}{\sqrt{\det\, \Big(S''(q_{-t})+iI\Big)\,\det\,\Phi''_{(\eta,\xi)}(q,p,q_{-t},p_{-t},t)}} \  .
\end{eqnarray}
One should note that in the leading order, the Hessian of the phase of the critical manifold, i.e. its restriction on the real critical manifold, ammounts to the pullback of the flow, $\Phi''_{(\eta,\xi)}(q,p,\eta,\xi,t)\equiv \Phi''_{(\eta,\xi)}(q,p,g^t(\eta,\xi))$, so that 
\begin{equation}
\Phi''_{(\eta,\xi)}(q,p,q_{-t},p_{-t},t)= \Phi''_{(\eta,\xi)}(q,p,q,p,0) \ , 
\end{equation}
since the time dependence of the anisotropy form is implicit in the flow, $Q(q,p,t)\equiv Q(g^{t}(q,p))$. We are lead to the result 
\begin{equation}
{\rm det}_{\mathbb{C}^{2d\times 2d}}\Phi''_{(\eta,\xi)}(q,p,q_{-t},p_{-t},t)={\rm det}_{\mathbb{C}^{d\times d}}(-I) \  .
\end{equation}
Then, according to the selection of the branch of the square root in subsection (B.3) of the appendix, we have
$$
\sqrt{\,\det\,\Phi''_{(\eta,\xi)}(q,p,q_{-t},p_{-t},t)} = (-i)^d
$$

Finally, we deduce that along the Lagrangian manifold $\Lambda_{S}^t$ the asymptotic solution is
\begin{equation}
\hspace{-25mm}\Psi_{wp}^\hbar(q,p,t)\sim (\pi\hbar)^{-d/4}\frac{i^{-3d/2}a(q_{-t})}{\sqrt{\det\, \Big(S''(q_{-t})+iI\Big)}}\exp\frac{i}{\hbar}\Big(S(q_{-t})+A(q_{-t},p_{-t},t)\Big) \  .
\label{eq:asympsol}
\end{equation}

We note that the transported phase constrained on the real critical manifold, 
\begin{equation}
S(q,t)=S(q_{-t})+A(q_{-t},p_{-t},t) \ ,
\end{equation}
where $q_{-t}=\underline q(-t;q,S'(q))$, is the solution of the corresponding \textit{Hamilton-Jacobi equation}
\begin{equation}
\frac{\partial S}{\partial t}+H\Big(q,\frac{\partial S}{\partial q}\Big)=0 \ , 
\end{equation}
with initial data $S(q,0)=S(q)$, so that $\Lambda_S^t=\Lambda_{S(\cdot, t)}$.

For reasons of convenience we have assumed above that the Lagrangian manifold $\Lambda^t_S$ is projectable, i.e., it can be diffeomorphically projected onto the base space. If this is not so, one can carry out the above arguments by introducing appropriate local coordinates, $(q_{i_1},\ldots,q_{i_m},p_{i_{m+1}},\ldots,p_{i_d})$, so that canonically conjugate coordinate pairs are excluded, as was introduced by Maslov in his construction of the canonical operator (e.g., \cite{lagrangian}). This overcomes the problem of caustics due to folds of the Lagrangian manifold $\Lambda^t_S$.

\subsection{Classical Limit of the Asymptotic Solution}

As the square modulus of the phase space wave function is a Husimi density (see subsection (\ref{wignr}) of the appendix), we expect that in the classical limit it will give a singular measure, concentrated on the Lagrangian manifold $\Lambda^t_S$.

For the asymptotic solution, we have 
\begin{eqnarray}
\hspace{-20mm}\int|\Psi^\hbar_{wp}(q,p,t)|^2\,dqdp\sim \int(\pi\hbar)^{-d/2}\,\frac{ 2^{d}\sqrt{\det\, \tilde Q_2}\,|\tilde a|^2\, e^{-\frac{2}{\hbar}{\rm Im}\,\tilde \Phi}}{|\det\,\Big((S''+iI)\,\tilde \Phi''_{(z,\zeta)}\Big)|}\, dqdp \ , 
\end{eqnarray}
which by the generalized stationary phase lemma \cite{generalized}, becomes to leading order
\begin{equation}
\int|\Psi^\hbar_{wp}(q,p,t)|^2\,dqdp\sim \int_{\Lambda_S^t}\frac{a^2}{|\det\,(S''+iI)|}\,d\nu \ .
\end{equation}

The induced Riemannian metric tensor on the transported Lagrangian manifold $\Lambda^t_S$ reads
\begin{equation}
g=\sum_{jl}\Bigg(\delta_{jl}+\Big(S''(q,t)^2\Big)_{jl}\Bigg)\,dq_jdq_l \ ,
\end{equation}
so that the Riemannian measure on the manifold is $d\nu=\sqrt{|g|}\,dq$, where 
\begin{equation}
\sqrt{|g|}=\sqrt{\det \Big(S''^2+I\Big)} \ .
\end{equation}

Having noted that $|\det\,(S''+iI)|^2=\det\,(S''^2+I)$, we deduce that 
\begin{equation}
\int|\Psi^\hbar_{wp}(q,p,t)|^2\,dqdp\sim\int a(x)^2\, dx =\|\psi^\hbar_0\|^2_{L^2(\mathbb{R}^d)}=1 \ ,
\end{equation}
in complete analogy to the WKBM asymptotic solution of the Schr\"{o}dinger equation. The difference, of course, is the existence of the Jacobian factor related to the solution of the transport equation (see, e.g., \cite{litt2}); lifted to the phase space, it is unity, the Jacobian of a symplectomorphism, modlulo the geometry of the transported Lagrangian manifold $\Lambda^t_S$.

\section{Two Simple Illustrations}

In order to put the semiclassical phase space propagator $\mathcal{K}_{wp}$ defined in $(\ref{eq:Lwp})$, to the test, we compare the asymptotic solutions it produces against solutions of the phase space Schr\"odinger equation for two simple idealized physical situations. In the first case we consider the propagation of a microparticle free of any external interactions, and in the other case scattering of a microparticle by a constant electric field, both in one dimension. 

In both cases we have subquadratic polynomial potentials,  $V(q)=\sum_{|\alpha|\leqslant 2}c_\alpha q^\alpha$, so they share the common characteristic of linearity of the Hamiltonian flow, and the phase space uniformity of the anisotropy form; the matrix Riccati equation is common for such systems, yielding the anisotropy form   
\begin{equation}
Z(t)=\frac{1}{2t-i} \ , \ \ W(t)=-\Big(Z(t)+i\Big)^{-1}=\frac{i}{2}\frac{1+2it}{1+it} \ ,
\end{equation}
constant in the initial phase space points. The phase space anisotropy form reads
\begin{equation}
 Q(t) = \frac{1}{2(1+it)}\left(
\begin{array}{rl}
i & -(1+2it) \\
-(1+2it) & i(1+2it)   
\end{array} \right) \ ,
\end{equation}
and 
\begin{equation}
\det\, Q_2(t)=\frac{1}{4(1+t^2)} \ , 
\end{equation}
while additionally, we have the phase 
\begin{equation}
\lambda(t)=-\frac{1}{2}{\rm arctan}\, (2t)+\frac{1}{2}{\rm arctan}\,\Big(\frac{t}{1+2t^2}\Big) \ .
\end{equation} 

We expect the semiclassical solution to be the exact solution, something characteristic for the class of quadratic potentials.

The quantum dynamics are given by the solution of the phase space initial value problem 
\begin{eqnarray}
\hspace{-20mm}i\hbar\frac{\partial\Psi}{\partial t}=-\hbar^2\frac{\partial^2\Psi}{\partial q^2}+V\Big(q+i\hbar\frac{\partial }{\partial p}\Big)\Psi \ , \ \ 0<t\leqslant T \nonumber\\
\hspace{-20mm}\Psi( \cdot ,0)=\Psi_0^\hbar=\mathcal{W}\psi_0^\hbar \in\mathfrak{F}^2\ ,
\end{eqnarray}
which stems from the configuration space problem,
\begin{eqnarray}
\hspace{-20mm}i\hbar\frac{\partial\psi}{\partial t}=-\hbar^2\frac{\partial^2\psi}{\partial x^2}+V(x)\psi \ , \ \ 0<t\leqslant T \nonumber\\
\hspace{-20mm}\psi( \cdot ,0)=\psi_0^\hbar \in L^2(\mathbb{R}^d) \ .
\end{eqnarray}

As initial data, for both cases, $V(x)=0$ and $V(x)=x$, we set off by considering appropriate semiclassical initial data for the Schr\"odinger equation, which allows for explicit calculations. We take semiclassical initial data, a complex phase WKBM state 
\begin{equation}
\psi^\hbar_0(x)=\pi^{-1/4}e^{-\frac{1}{2}x^2}e^{\frac{i}{2\hbar}x^2} \ ,
\end{equation}
so that the initial amplitude and phase are, respectively, $a(x)=\pi^{-1/4}e^{-x^2/2}$ and $S(x)=\frac{1}{2}x^2$, generating the Lagrangian manifold $\Lambda_{S}=\Big\{(q,p):\, p=q\Big\}$. 

The phase space image of the initial data $\psi_0^\hbar$ under the wave packet transform is 
\begin{equation}
\hspace{-20mm}\Psi^\hbar_0(q,p):=\mathcal{W}\psi_0^\hbar(q,p)=\hbar^{-1/4}\sqrt{\frac{1}{\pi}\frac{1}{1-i+\hbar}}e^{-p^2/2\hbar} \exp\frac{1}{2\hbar}\frac{i-\hbar}{1-i+\hbar}z^2 \ ,
\end{equation}
for $z=q-ip$.

\subsection{Free Motion}
The Hamiltonian is
\begin{equation}
H(q,p)=p^2 \ ,
\end{equation}
which induces the flow $g^t(q,p)=(q+2tp,p)$ , $t\geqslant 0$. The phase space action along the flow is $A(q,p,t)=p^2t$.

The solution is obtained by means of the free propagator \cite{Feynman},
\begin{eqnarray}
\hspace{-20mm}\psi^\hbar(x,t)=\frac{1}{(4\pi i \hbar t)^{1/2}}\int e^{\frac{i}{4\hbar t}(x-y)^2}\psi_0^\hbar(y)\,dy \ .
\end{eqnarray}

For the particular initial data, we obtain,
\begin{equation}
\hspace{-20mm}\psi^\hbar(x,t)=\pi^{-1/4}\sqrt{\frac{1}{1+2(1+i\hbar)t}}\exp\frac{i}{2\hbar}\Big(\frac{1+i\hbar}{1+2(1+i\hbar)t}x^2\Big) \ ,
\end{equation}
which in phase space, becomes, by applying the wave packet transform,
\begin{eqnarray}
\hspace{-20mm}\Psi^\hbar(q,p,t)=\hbar^{-1/4}\sqrt{\frac{1}{\pi}\frac{1}{1-i+\hbar+2(1+i\hbar)t}}\nonumber \\
\hspace{-20mm}\times \exp\frac{i}{2\hbar}\frac{-2i(1+i\hbar)qp+(1+i\hbar)q^2+i(1+2(1+i\hbar)t)p^2}{1-i+\hbar+2(1+i\hbar)t} \ .
\end{eqnarray}

The semiclassical wave packet propagator is 
\begin{eqnarray}
\hspace{-20mm}\mathcal{K}_{wp}(q,p,\eta,\xi,t;\hbar)=\frac{e^{-\frac{i}{2}\Big({\rm arctan}\, (2t)+{\rm arctan}\,\Big(\frac{1+2t^2}{t}\Big)+\frac{\pi}{2}\Big)}}{2\pi\hbar(1+t^2)^{1/4}}\exp\frac{i}{\hbar}\bigg(\xi^2 t-p(\eta-q+2\xi t)\nonumber \\
+\frac{1}{4(t-i)}\Big[(q-\eta)^2-2i(q-\eta)\xi+(1+2it)p^2\nonumber \\
+(1+6it-4t^2)\xi+2p(i-2t)(q-\eta+i\xi-2t\xi)\Big]\bigg) \ .
\end{eqnarray}

The semiclassical solution by the semiclassical wave packet propagator, is given by 
\begin{equation}
\Psi^\hbar_{wp}(q,p,t)=\int\mathcal{K}_{wp}(q,p,\eta,\xi,t;\hbar)\Psi^\hbar_0(\eta,\xi)\, d\eta d\xi \ .
\end{equation}

For the particular choice of initial data above, the asymptotic solution constructed by means of the semiclassical wave packet propagator, $\mathcal{U}^t_{wp}$,
\begin{eqnarray}
\hspace{-20mm}\Psi_{wp}^\hbar(q,p,t)=\hbar^{-1/4}\sqrt{\frac{1}{\pi}\frac{1}{1-i+\hbar+2(1+i\hbar)t}}\nonumber \\
\hspace{-20mm}\times\sqrt{\frac{1+it}{\sqrt{1+t^2}}}e^{-\frac{i}{2}({\rm arctan}\, (2t)-{\rm arctan}\,\Big(\frac{t^2+1}{t}\Big))}\nonumber \\
\hspace{-20mm}\times \exp\frac{i}{2\hbar}\frac{-2i(1+i\hbar)qp+(1+i\hbar)q^2+i(1+2(1+i\hbar)t)p^2}{1-i+\hbar+2(1+i\hbar)t} \ .
\end{eqnarray}

We unsurprisingly note that the asymptotic solution bore by the wave packet propagator, is actually identified with the exact solution.

The asymptotic solution, and thus the solution itself, is semiclassically concentrated on the Lagrangian manifold as transported by Hamilton-Jacobi dynamics. In particular, the Hamilton-Jacobi problem 
\begin{eqnarray}
\frac{\partial S}{\partial t}+\Big(\frac{\partial S}{\partial x}\Big)^2=0 \ , \ \ t\in\mathbb{R}_+\nonumber\\
S(x,0)=S(x)=\frac{1}{2}x^2 \ ,
\end{eqnarray}
has the solution 
\begin{equation}
S(x,t)=\frac{1}{2}\frac{x^2}{1+2t} \ ,
\end{equation}
which induces the transported Lagrangian manifold 
\begin{equation} 
\hspace{-20mm}\Lambda^t_{S}=\Big\{(q,p):\,p=\partial_q S(q,t)\Big\}=\Big\{(q,p):\,p=\frac{q}{1+2t}\Big\} \ ,
\end{equation}
i.e., initially the diagonal straight line $\Lambda_{S}=\Big\{(q,p):\, p=q\Big\}$, asymptotically tending to the horizontal straight line $p=0$.

\subsection{Scattering by a Constant Electric Field}
The Hamiltonian is
\begin{equation}
H(q,p)=p^2+q \ ,
\end{equation}
which induces the flow $g^t(q,p)=(q+2tp-t^2,p-t)$, $t\geqslant 0$. The phase space action along the flow is $A(q,p,t)=(p^2-q)t-2pt^2+\frac{2}{3}t^3$.

The solution of the phase space problem is obtained by means of the Airy propagator \cite{Airy}, 
\begin{eqnarray}
\psi^\hbar(x,t)=\frac{e^{-\frac{i}{\hbar}\Big(\frac{1}{3}t^3+tx\Big)}}{2(2\pi i\hbar)^{1/2}}\int e^{-\frac{1}{4i\hbar t}(x-y+t^2)}\psi_0^\hbar(y)\,dy \ .
\end{eqnarray}

For the particular initial data, we obtain,
\begin{eqnarray}
\hspace{-20mm}\psi^\hbar(x,t)=\pi^{-1/4}\sqrt{\frac{1}{1+2(1+i\hbar)t}}\nonumber \\
\times \exp\frac{i}{\hbar}\Bigg(\frac{1}{2}\frac{(t^2+x)^2}{1+2(1+i\hbar)t}-\frac{1}{3}t^3-tx\Bigg) \exp-\frac{1}{2}\frac{(t^2+x)^2}{1+2(1+i\hbar)t} \ ,
\end{eqnarray}
which in phase space, becomes, by applying the wave packet transform,
\begin{eqnarray}
\hspace{-20mm}\Psi^\hbar(q,p,t)=\hbar^{-1/4}\sqrt{\frac{1}{\pi}\frac{1}{1-i+\hbar+2(1+i\hbar)t}}\nonumber \\
\hspace{-20mm}\times \exp\frac{i}{\hbar}\frac{1}{1+2(1+i\hbar)t}\Bigg(\frac{i}{2}\frac{\Big[i(1+2(1+i\hbar)t)(q-ip)+(1+(1+i\hbar)t)t\Big]^2}{1-i+\hbar+2(1+i\hbar)t}\nonumber \\
+\frac{i}{2}(q^2-2iqp)(1+2(1+i\hbar)t)-\frac{1}{3}t^3\bigg[1+\frac{1}{2}(1+i\hbar)t\bigg]\Bigg) \ .
\end{eqnarray}

The semiclassical wave packet propagator is 
\begin{eqnarray}
\hspace{-20mm}\mathcal{K}_{wp}(q,p,\eta,\xi,t;\hbar)=\frac{e^{-\frac{i}{2}\Big({\rm arctan}\, (2t)+{\rm arctan}\,\Big(\frac{1+2t^2}{t}\Big)+\frac{\pi}{2}\Big)}}{2\pi\hbar(1+t^2)^{1/4}}\nonumber \\
\times \exp\frac{i}{\hbar}\Big(\frac{2}{3}t^3+(p-2\xi)t^2+(\xi^2-\eta-2p\xi)t+p(q-\eta)\Big)\nonumber \\
\hspace{-20mm}\times \exp\frac{i}{4\hbar(1+t^2)}\bigg(t(q-\eta_t)^2-t(p-\xi_t)^2-2(1+2t^2)(q-\eta_t)(p-\xi_t)\nonumber \\
\hspace{-20mm}+i\Big[(q-\eta_t)^2+(1+2t)(p-\xi_t)^2-2t(q-\eta_t)(p-\xi_t)\Big]\bigg) \ .
\end{eqnarray}

And so, the semiclassical solution by the semiclassical wave packet propagator, is given by 
\begin{equation}
\Psi^\hbar_{wp}(q,p,t)=\int\mathcal{K}_{wp}(q,p,\eta,\xi,t;\hbar)\Psi^\hbar_0(\eta,\xi)\, d\eta d\xi \ .
\end{equation}
For the particular choice of initial data above, we have,
\begin{eqnarray}
\hspace{-25mm}\Psi^\hbar_{wp}(q,p,t)=\hbar^{-1/4}\sqrt{\frac{1}{\pi}\frac{1}{1-i+\hbar+2(1+i\hbar)t}}\nonumber \\
\hspace{-25mm}\times\sqrt{\frac{1+it}{\sqrt{1+t^2}}}e^{-\frac{i}{2}({\rm arctan}\, (2t)-{\rm arctan}\,\Big(\frac{t^2+1}{t}\Big))}\nonumber \\
\hspace{-25mm}\times \exp\frac{i}{2\hbar}\Bigg(it^2+\frac{2i}{3}(i+1+i\hbar)t^3-\frac{1}{3}(1+i\hbar)t^4\nonumber \\
\hspace{-25mm}-2t\Big[1+(1+i\hbar)t\Big](q-ip)+(1+i\hbar)q^2+\Big[1+2(1+i\hbar)\Big]tp^2-2i(1+i\hbar)qp\Bigg) \ .
\end{eqnarray}
As in the case of free motion, the asymptotic solution is identified with the exact solution.

The asymptotic solution, and thus the solution itself, is semiclassically concentrated on the Lagrangian manifold as transported by Hamilton-Jacobi dynamics. In particular, the Hamilton-Jacobi problem 
\begin{eqnarray}
\frac{\partial S}{\partial t}+\Big(\frac{\partial S}{\partial x}\Big)^2+x=0 \ , \ \ t\in\mathbb{R}_+ \nonumber\\
S(x,0)=S(x)=\frac{1}{2}x^2 \ ,
\end{eqnarray}
has the solution 
\begin{equation}
S(x,t)=\frac{1}{2(1+2t)}\Big(x^2-2t(1+t)x-\frac{1}{3}(2+t)t^3\Big) \ ,
\end{equation}
which induces the transported Lagrangian manifold 
\begin{equation} 
\hspace{-20mm}\Lambda^t_{S}=\Big\{(q,p):\,p=\partial_q S(q,t)\Big\}=\Big\{(q,p):\,p=\frac{q-t(t+1)}{1+2t}\Big\} \ ,
\end{equation}
i.e., initially the diagonal straight line $\Lambda_{S}=\Big\{(q,p):\, p=q\Big\}$, asymptotically tending to the horizontal straight line $p=-\frac{t}{2}$.

\section{Discussion}
We have considered the singular initial value problem for the Schr\"{o}dinger equation in phase space and its semiclassical asymptotic solutions for WKBM phase space initial data. We pass to the phase space representation by conjugating with the wave packet transform. We introduce the phase space quantum flow, and the phase space propagator, and give a well posed initial value problem in the phase space. 

The asymptotic solutions are constructed via a semiclassical propagator for the phase space Schr\"{o}dinger equation, the wave packet propagator, based on the anisotropic Gaussian approximation.

The wave packet propagator defines a linear, unitary quantum flow on the Fock-Bargmann space, meaning that semiclassical quantum evolution preserves an initially pure state as such. 

It is underlied by a rigid algebraic structure, characteristic of the anisotropic Gaussian approximation and the nearby orbit approximation on which it is constructed. It closely resembles the wave packet quantization of a canonical transformation \cite{naz1,naz2}.

The wave packet propagator is seen in a weak limit to approach the Gutzwiller-van Vleck propagator. It is also tested against complex phase WKBM initial data, reproducing the exact solutions for the phase space initial value problem in the simple situations of free motion and scattering off a constant electric field. 

The asymptotic solution produced by the semiclassical wave packet propagator yields a semiclassical Husimi function on the Lagrangian manifold $\Lambda_S^t$, where $S$ is the initial WKBM phase; alternatively, the Lagrangian manifold generated by the solution of the Hamilton-Jacobi equation with initial data $S(q)$.

The relation to the Maslov canonical operator can be seen by taking the inverse wave packet transform of the asymptotic solution $(\ref{eq:asympsol})$, 
\begin{equation}
\psi^\hbar(x,t)\sim \Big(\frac{1}{2\pi\hbar}\Big)^{d/2}\int G_{(q,p)}(x;\hbar)\Psi^\hbar_{wp}(q,p,t)\, dqdp \ ,
\end{equation}
becoming asymptotically an integral over $\Lambda^t_S$.

The authors would like to thank Sergey Dobrokhotov, Frederic Faure, Maurice de Gosson, Robert Littlejohn, Vladimir Nazaikinskii, Vesselin Petkov and Roman Schubert for fruitfull conversations and enlightening comments.

\appendix
\renewcommand*{\thesection}{\Alph{section}}

\section{The Wave Packet Transform and Quantization}

\subsection{The Wave Packet Transform}
\label{twpt}
The wave packet transform defines a bounded operator 
\begin{equation}
\mathcal{W}:L^2(\mathbb{R}^d,\mathbb{C};dx)\rightarrow \mathfrak{F}^2 \ ,  
\end{equation}
whose image space, \textit{the Fock-Bargmann space}, is a subspace of $L^2(\mathbb{R}^{2d},\mathbb{C};dqdp)$ \cite{naz2}. It satisfies the Plancherel equality 
\begin{equation}
\int\bar \psi\varphi \,dx=\int\overline{\mathcal{W}\psi}\, \mathcal{W}\varphi \, dqdp \ ,
\end{equation}
rendering it an isometric isomorphism, which guarantees the unitarity of the quantum flow in phase space and the conservation of the phase space probability \cite{naz2}. 

The wave function $\psi$ is represented in phase space by analyzing it in the overcomplete set of isotropic Gaussian wave packets, $\{G_{(q,p)}\}_{(q,p)\in\mathbb{R}^{2d}}$.
Isotropic Gaussian wave packets constitute a nondenumerable, overcomplete basis in $L^2(\mathbb{R}^d)$, in the sense that any $\psi\in L^2(\mathbb{R}^d)$ can be written uniquely as a superposition of coherent states, which are, however, not orthogonal \cite{naz2,rober1,naz1},
\begin{equation}
\hspace{-20mm}\langle G_{(q,p)},G_{(\eta,\xi)}\rangle_{L^2(\mathbb{R}^d)}=(\pi\hbar)^{d/2}e^{\frac{i}{\hbar}\Big(p\cdot q-\eta\cdot \xi+\frac{1}{2}(\eta+q)\cdot(\xi-p)+\frac{i}{4}(q-\eta)^2+\frac{i}{4}(p-\xi)^2\Big)} \ .
\end{equation}
Rather, they are asymptotically orthogonal, 
\begin{equation}
\langle G_{(q,p)},G_{(\eta,\xi)}\rangle_{L^2(\mathbb{R}^d)}=O(\hbar^\infty) \ ,
\end{equation}
for fixed $(q,p)\neq (\eta,\xi)$, or, more specifically,
\begin{equation}
\bigg(\frac{1}{\pi\hbar}\bigg)^{d/2}\langle G_{(q,p)},G_{(\eta,\xi)}\rangle_{L^2(\mathbb{R}^d)}\rightharpoonup_*\delta(q-\eta)   \delta(p-\xi) \ .
\end{equation} 

While they fail to satisfy a completeness relation in $L^2(\mathbb{R}^d)$, they satisfy a completeness relation in phase space \cite{naz2,rober1},  
\begin{equation}
\bigg(\frac{1}{2\pi\hbar}\bigg)^{d}  \int\bar G_{(q,p)}(x;\hbar)G_{(q,p)}(y;\hbar)\,dqdp=\delta(x-y) \ .
\end{equation}

The main physical significance of coherent states is that they are the optimal wave functions in minimizing the product of the position-momentum uncertainties, naturally associating them with pure mechanical states, i.e., points in phase space. If $\hat x=m_x$ is the multiplication operator by $x$ and $\hat p_x=-i\hbar \partial_x$, it is straightforward to incur the uncertainty relations while in the state $G_{(q,p)}$,
\begin{equation}
\Delta\hat x_j = \Delta \hat p_{x_j} =\sqrt{\frac{\hbar}{2}} \ ,
\end{equation} 
so that there is a saturation of the Heisenberg inequality, $\Delta \hat x_j \, \Delta \hat p_{x_j}=\frac{\hbar}{2}$ \cite{naz1}.

The phase space wave function, $\Psi(q,p)=\mathcal{W}\psi(q,p)$, allows a statistical interpretation: \textit{is the probability amplitude of a microparticle being in the quantum coherent state} $G_{(q,p)}(x)$. It cannot be interpreted in an analogous way to Schr\"{o}dinger wavefunctions, i.e., the probability amplitude of a microparticle's position and momentum being $(q,p)$, as the maximum resolution in phase space is the Planck cell. This reflects the fact that in phase space quantum theory the pure state is not related to a quantum state supported in a sense at that point, but a coherent state microlocalized on that point on the Heisenberg scale.
An equivalent way of defining the wave packet transform is by \textit{the  Weyl operator} \cite{litt1,cohstates}, 
\begin{equation}
\mathcal{T}_z:=e^{\frac{i}{\hbar}\Big(p\cdot \hat x-q\cdot \hat p_x\Big)}=e^{\frac{i}{\hbar}\Big(p\cdot x+i\hbar q\cdot \partial_x\Big)} \ ,
\end{equation}
where $z=q-ip$. The Weyl operator is unitary, and has the following property 
\begin{equation}
\langle \psi,\hat z\psi\rangle_{L^2(\mathbb{R}^d)}=z\,\Rightarrow \, \langle \psi,\mathcal{T}^*_\zeta\circ \hat z\circ \mathcal{T}_\zeta\psi\rangle_{L^2(\mathbb{R}^d)}=z-\zeta \ ,
\end{equation}
for $\zeta=\eta-i\xi$, as well as the group structure,
\begin{equation}
\mathcal{T}_z\circ \mathcal{T}_\zeta=\mathcal{T}_{z+\zeta} \ , \ \ \mathcal{T}_z^{-1}=\mathcal{T}^*_z=\mathcal{T}_{-z} \ , \ \ \mathcal{T}_0={\rm Id}_{L^2(\mathbb{R}^d)} \ .
\end{equation}
By the Baker-Cambell-Hausdorff formula, it is simplified to $\mathcal{T}_z=e^{-\frac{i}{2\hbar}p\cdot q}e^{\frac{i}{\hbar}p\cdot x}e^{-q\cdot \partial_x}$.

The Wey operator generates the set of isotropic Gaussian wave packet by its action on the vacuum Gaussian state, $G_0(x;\hbar)=(\pi\hbar)^{-d/4}e^{-x^2/2\hbar}$, 
\begin{equation}
\mathcal{T}_zG_0(x)=G_{(q,p)}(x;\hbar) \ .
\end{equation}
Its action on general Gaussian wave packets is simply that of phase space translation,
\begin{equation}
\mathcal{T}_z\mathcal{G}^Z_\kappa(x;\hbar)=\mathcal{T}_z(\pi\hbar)^{-d/4}e^{\frac{i}{\hbar}\Big(\kappa\cdot x+\frac{1}{2}x\cdot Zx\Big)}=\mathcal{G}^Z_{\kappa-p}(x-q;\hbar) \ .
\end{equation}
By means of the Weyl operator, the wave packet transform can thus be expressed
\begin{equation}
\mathcal{W}\psi(q,p)=\bigg(\frac{1}{2\pi\hbar}\bigg)^{d/2}{\rm tr}\Big(\mathcal{T}_z^*\psi\langle G_0,\cdot \rangle_{L^2(\mathbb{R}^d)}\Big)(q,p) \ .
\end{equation}

\subsection{The Analytic Structure of the Wave Packet Transform}
\label{analytic}

The following commutation relations of the wave packet transform with the conjugate canonical dynamical variables $\hat x$ and $\hat p_x$ hold \cite{torres,harriman,naz2}:
\begin{equation}
\hspace{-20mm}\mathcal{W} \circ \hat x=(q+i\hbar \partial_p)\circ \mathcal{W} \ \ \ {\rm and} \ \ \ \mathcal{W}\circ (-i\hbar \partial_x)=(-i\hbar \partial_q)\circ  \mathcal{W}  \ .
\label{eq:rels}
\end{equation}
These relations define the wave packet image of the conjugate canonical pair, namely
\begin{equation}
\check q=q+i\hbar \frac{\partial}{\partial p} \ , \ \ \check p=-i\hbar\frac{\partial }{\partial q} \ .
\end{equation} 
Rearranging the above commutation relations, one readily incurs that any $\Psi= \mathcal{W} \psi\in\mathfrak{F}^2$ is constrained by the relation
\begin{equation}
\check {\bar z}\,\Psi=\bar z\,\Psi \ ,
\end{equation}
where $\bar z=q+ip$, and $\check{\bar z}=\check{z}^*$ is the annihilation operator, in terms of many body quantum mechanics. Such a constraint on phase space wave functions comes as little surprise, as the doubling of the dimensionality of the base space introduces a great redundancy in the image space. 

In terms of the above constraint, \textit{the Fock-Bargmann constraint} \cite{torres,polish}, the Fock-Bargmann space $\mathfrak{F}^2$, as a subspace of $L^2(\mathbb{R}^{2d})$ can be written as
\begin{equation} 
\hspace{-20mm}\mathfrak{F}^2=\bigcap_j{\rm ker}\Big(\hbar\partial _{q_j}-i\hbar\partial_{ p_j}-ip_j \Big)=\bigcap_j{\rm ker}\Big(\partial_{q_j}-i\partial_{p_j}\Big)\Big(e^{p^2/2\hbar}(\cdot)\Big) \ ,
\label{eq:ker1}
\end{equation}
making explicit its Gaussian `twisted' analyticity property.

The Fock-Bargmann constraint defines an analyticity condition on the Fock-Bargmann space and the projector onto $\mathfrak{F}^2$, which is elucidated by considering \textit{the complex Fock-Bargmann space}\footnote{This is actually the Hilbert space of analytic functions discovered by Bargmann \cite{bargmann1,bargmann2}}, $\mathfrak{F}^2_{\mathbb{C}}$, which comprizes of analytic functions on $\mathbb{C}^d$, embedded in complex Gaussian weighted $L^2$ space, $\mathfrak{F}^2_{\mathbb{C}}\subset L^2(\mathbb{C}^d,\mathbb{C};d\mu)$, where
\begin{equation}
d\mu(z,\bar z)=\frac{e^{-\bar z\cdot z}}{\pi^d}\, dm(z,\bar z)=\Big(\frac{2i}{\pi}\Big)^de^{-\bar z\cdot z} \,d\bar zdz
\end{equation}
is the Gaussian, and $dm$ the Lebesgue measures on $\mathbb{C}^d$ respectively. Here we have \textit{microlocalized} symplectic complex pseudocoordinates, on the Heisenberg scale, meaning 
\begin{equation}
z=\frac{q-ip}{\sqrt{2\hbar}} \ , \ \ \bar z=\frac{q+ip}{\sqrt{2\hbar}} \ ,
\label{eq:ta1}
\end{equation}
reserving the same notation $z,\bar z$ for convenience. The above is made quite transparent by the form $(\ref{eq:twist})$ of the Fock-Bargmann constraint.

The isomorphism between the Fock-Bargmann spaces explicitly reads 
\begin{equation}
\chi:\mathfrak{F}^2\rightarrow \mathfrak{F}^2_{\mathbb{C}} \, \Big| \, \Psi\mapsto f=\chi \Psi \ ,
\label{eq:ta2}
\end{equation}
where \cite{naz2}
\begin{eqnarray}
\chi\Psi(z)= g(z,\bar z)\Psi\circ M^{-1}(z,\bar z)=:f(z) \nonumber\\ 
\chi^{-1}f(q,p)=\frac{f\circ M(q,p)}{ g\circ M(q,p)}=\Psi(q,p) \ ,
\label{eq:ta3}
\end{eqnarray} 
with $M$ the linear symplectic phase space pseudocomplexification,
\begin{equation}
M=\frac{1}{\sqrt{2\hbar}}\left(
\begin{array}{ccc}
I & -iI \\
I & iI 
\end{array} \right) \ ,
\label{eq:ta4}
\end{equation}
the overall factor $g$ being a Gaussian, 
\begin{equation}
g(z,\bar z)=(2\pi \sqrt{\hbar})^{-d/2}e^{\frac{1}{4}(z-\bar z)^2-\frac{1}{2}z^2} \ .
\label{eq:ta5}
\end{equation}
The above isomorphism maps the Fock-Bargmann constraint on phase space wave functions to the Cauchy-Riemann analyticity conditions, $\bar \partial_z f=0$, so that as a subspace of $L^2(\mathbb{C}^d)$, the complex Fock-Bargmann space is 
\begin{equation}
\mathfrak{F}^2_{\mathbb{C}}= \bigcap_j{\rm ker}\,\bar \partial_{z_j} \ .
\label{eq:ta6}
\end{equation}

The twisted analyticity properties of phase space wave functions, and their connection to the space of Gaussian weighted phase space analytic functions, stems from the connection of the wave packet transform to \textit{the Bargmann transform} \cite{bargmann1,bargmann2}. The wave packet transform is a `semiclassical microlocalization' of the Bargmann transform. Explicitly, the analytic function determined by the isomorphism $\chi$ is the Bargmann transform of $\psi$ localized on the Heisenberg scale,
\begin{equation}
f(z)=\mathcal{B}u(z)=\frac{1}{\pi^{d/4}}\int e^{-\frac{1}{2}(z^2+x^2)+\sqrt{2}z\cdot x}u(x)\,dx \ .
\end{equation}

The induced topological structures of the two spaces are given by the inner products
\begin{equation}
\langle \Psi,\Xi\rangle_{\mathfrak{F}^2}:=\int \bar \Psi\,\Xi \, dqdp \ , \ \|\Psi\|_{\mathfrak{F}^2}^2:=\langle\Psi,\Psi\rangle_{\mathfrak{F}^2}
\end{equation}
and 
\begin{equation}
\langle f,g\rangle_{\mathfrak{F}^2_{\mathbb{C}}}:=\int\bar f\,g\,d\mu(z,\bar z) \ , \ \ \|f\|_{\mathfrak{F}^2_{\mathbb{C}}}^2:=\langle f,f\rangle_{\mathfrak{F}^2_{\mathbb{C}}} \ .
\end{equation}

The adjoint operator of $\mathcal{W}$ is \cite{naz2}
\begin{equation}
\mathcal{W} ^*: L^2(\mathbb{R}^{2d})\rightarrow L^2(\mathbb{R}^{d})\,\Big| \ \ \Psi\mapsto  \mathcal{W} ^*\Psi \ ,
\end{equation}
where 
\begin{eqnarray}
\hspace{-20mm}\mathcal{W}^*\Psi(x)=\bigg(\frac{1}{2\pi\hbar}\bigg)^{d/2}\lim_{n}\int G_{(q,p)}(x;\hbar)\Psi_n(q,p)\,dqdp\nonumber \\
=:\bigg(\frac{1}{2\pi\hbar}\bigg)^{d/2}\int^*G_{(q,p)}(x;\hbar)\Psi(q,p)\,dqdp \ ,
\label{eq:star}
\end{eqnarray}
where $\{\Psi_n\}$ is a sequence of bounded, compactly supported phase space functions with $\Psi_n\rightarrow \Psi$ strongly in $L^2(\mathbb{R}^{2d})$. The adjoint enables us to introduce the orthogonal projection $\mathcal{P}_{\mathfrak{F}^2}:L^2(\mathbb{R}^{2d})\rightarrow\mathfrak{F}^2$ \cite{naz2},
\begin{equation}
\mathcal{P}_{\mathfrak{F}^2}:=\mathcal{W}\circ \mathcal{W} ^* \ ,
\label{eq:projector1}
\end{equation} 
which plays a central role in the wave packet quantization formalism. Restricted on $\mathfrak{F}^2$, the adjoint $\mathcal{W}^*$ is identified with the inverse $\mathcal{W}^{-1}$ \cite{naz2},
\begin{equation}
\mathcal{W} ^* |_{\mathfrak{F}^2}= \mathcal{W} ^{-1} \ \ {\rm and} \ \ \mathcal{W}\circ  \mathcal{W} ^* |_{\mathfrak{F}^2} ={\rm Id}_{\mathfrak{F}^2} \ .
\label{eq:projector2}
\end{equation}
In particular, for $\Psi\in\mathfrak{F}^2$
\begin{eqnarray}
\hspace{-20mm} \mathcal{W} ^{-1}\Psi^\hbar(x)=\bigg(\frac{1}{2\pi\hbar}\bigg)^{d/2}\int^* G_{(q,p)}(x;\hbar)\Psi^\hbar(q,p)\,dqdp\nonumber \\
=\bigg(\frac{1}{2\pi\hbar}\bigg)^{d/2}\int G_{(q,p)}(x;\hbar)\Psi^\hbar(q,p)\,dqdp \ .
\end{eqnarray}

There is a simple physical interpretation of the Fock-Bargmann decomposition of the phase space Hilbert space, 
\begin{equation}
L^2(\mathbb{R}^{2d})=\mathfrak{F}^2\oplus (\mathfrak{F}^2)^{\perp} \ ,
\end{equation} 
in phase space functions which are images of Schr\"{o}dinger wave functions, and those which are not. 
The $L^2$ phase space functions outside $\mathfrak{F}^2$ account for \textit{mixed quantum states}, which do not correspond to Schr\"{o}dinger wave functions, while the elements of $\mathfrak{F}^2$ account for \textit{pure quantum states}.

The Fock-Bargmann projector $\mathcal{P}_{\mathfrak{F}^2}$ induces equivalence classes of mixed quantum states $[\Psi]\subset L^2(\mathbb{R}^{2d})$ equivalent to the corresponding pure state $\Psi_0=\mathcal{P}_{\mathfrak{F}^2}\Psi$, modulo scaling. The local properties of $\mathcal{P}_{\mathfrak{F}^2}$ are a smoothing on the Heisenberg scale, smearing away oscillations on that scale 
\begin{equation}
\hspace{-20mm}\mathcal{P}_{\mathfrak{F}^2}\Psi(q,p)=\bigg(\frac{1}{2\pi\hbar}\bigg)^{d} \int\!\!\!\int^*e^{\frac{i}{\hbar}\Big(p\cdot (y-x)+\frac{i}{2}\Big[(x-q)^2+(y-q)^2\Big]\Big)}\Psi (\eta,\xi)\,dxd\eta d\xi \ .
\label{eq:Berg1}
\end{equation}

For $\mathfrak{F}^2_\mathbb{C}$, there is a smooth kernel analogue of the Dirac distribution customary in nonanalytic Hilbert spaces. In particular, the kernel 
\begin{equation}
B_{\mathfrak{F}^2_\mathbb{C}}(z,\bar \zeta):=e^{z\cdot \bar \zeta} \ ,
\label{eq:Berg2}
\end{equation}
\textit{the Bergmann reproducing kernel}, has the reproducing property \cite{bargmann1,bargmann2}
\begin{equation}
f(z)=\int B_{\mathfrak{F}^2_\mathbb{C}}(z,\bar\zeta)f(\zeta)\,d\mu(\zeta,\bar\zeta ) \ .
\label{eq:Berg3}
\end{equation}
It is the integral kernel of the pseudocomlplexified Fock-Bargmann projector, $f(z)=\mathcal{P}_{\mathfrak{F}^2_{\mathbb{C}}}f(z)$, where 
\begin{equation}
\mathcal{P}_{\mathfrak{F}^2_{\mathbb{C}}}=\chi\circ \mathcal{P}_{\mathfrak{F}^2}\circ \chi^{-1} \ .
\end{equation}

\subsection{The Wave Packet Quantization}
\label{pqs}

In this subsection we make note of the relation between the Heisenberg to the wave packet quantization. We reserve the notations $f\mapsto {\rm Op}_{H}(f)$ and $f\mapsto {\rm Op}_{wp}(f)$ for the two quantizations respectively. 

\textit{The Heisenberg quantization of} $f$ is defined as 
\begin{equation}
{\rm Op}_{H}(f):=\mathcal{F}^{*}\circ m_f\circ \mathcal{F}=f\Big(\stackrel{2}{x},-i\hbar \stackrel{1}{\partial _x}\Big) \ , 
\label{eq:H}
\label{eq:Toplitz}
\end{equation}
where $m_f$ is the multiplication operator by $f$, and $\mathcal{F}$ is the semiclassical Fourier transform, $\mathcal{F} f(\xi)=\Big(\frac{1}{2\pi\hbar}\Big)^{d/2}\int e^{-\frac{i}{\hbar}x\cdot \xi}f(x)\,dx$. As a pseudodifferential operator, it reads
\begin{equation}
{\rm Op}_{H}(f)\psi^\hbar(x)=\Big(\frac{1}{2\pi\hbar}\Big)^d\int e^{\frac{i}{\hbar}p\cdot (x-q)}f(x,p)\psi^\hbar(q)\, dqdp \ ,
\end{equation}
while for analytic $f$, we have the representation
\begin{equation}
\hspace{-25mm} {\rm Op}_{H}(f)\psi^\hbar(x)=f\Big(\stackrel{2}{x},-i\hbar\stackrel{1}{\partial}_x\Big)\psi^\hbar(x)=\sum_{\alpha,\beta\in\mathbb{Z}^d_+}\frac{\partial^{\alpha}_x\partial^{\beta}_{p_x}f(0)}{\alpha!\beta!}x^{\alpha}(-i\hbar\partial _x)^{\beta}\psi^\hbar(x) \ .
\end{equation}

\textit{The wave packet quantization of} $f$ is defined according to the rule \cite{naz2}
\begin{equation}
\langle \psi^\hbar,{\rm Op}_{wp}(f)\psi^\hbar\rangle _{L^2(\mathbb{R}^d)}=\langle \mathcal{W}\psi^\hbar,f \mathcal{W}\psi^\hbar\rangle_{\mathfrak{F}^2} \ .
\end{equation}
In particular, we have 
\begin{equation}
{\rm Op}_{wp}(f):=\mathcal{W}^*\circ m_f\circ\mathcal{W}= f\Big(\stackrel{2}{x},-i\hbar \stackrel{1}{\partial _x}\Big)+ O(\hbar^\infty) \ , 
\end{equation}
a definition analogous to $(\ref{eq:H})$. The difference can be understood in noting the rigidity in the order of integration in a Fourier inegral operator representation of the later, 
\begin{equation}
\hspace{-20mm}{\rm Op}_{wp}(f)\psi^\hbar(x)=\Big(\frac{1}{2\pi\hbar}\Big)^d\int\!\!\!\int^*G_{(\eta,\xi)}(x;\hbar)\bar G_{(\eta,\xi)}(y;\hbar)f(\eta,\xi)\psi^\hbar(y)\,dyd\eta d\xi \ ,
\end{equation}
according to definition $(\ref{eq:star})$.

The operator ${\rm Op}_{wp}(f)$ has the alternative form 
\begin{equation}
{\rm Op}_{wp}(f)=\mathcal{W}^{-1}\circ \mathcal{P}_{\mathfrak{F}^2}\circ m_f\circ \mathcal{W} \ ,
\end{equation}
where $\mathcal{P}_{\mathfrak{F}^2}=\mathcal{W}\circ \mathcal{W}^*$ is the projector onto the Fock-Bargmann space $\mathfrak{F}^2$. 

As they ought to, the above quantization rules share a common weak classical limit. If ${\rm Op}_{wp}(f)=g\Big(\stackrel{2}{x},-i\hbar\stackrel{1}{\partial}_x;\hbar\Big)$, then we have the semiclassical relation \cite{naz2}
\begin{equation}
g(q,p;\hbar)\sim \sum_{k=0}^{\infty}\frac{\hbar^k}{k!}\Bigg(\frac{1}{4}\Delta_{(q,p)}-\frac{i}{2}\partial_q\cdot \partial_p\Bigg)^kf(q,p) \ ,
\end{equation}
differing, in general, even for polynomial symbols.

These operators can be expressed in the phase space representation, by means of conjugations with the wave packet transform. One should, however, take care not to confuse \textit{the wave packet quantization of some physical quantity} $f$, which is an operator on an appropriate subspace of $L^2(\mathbb{R}^d)$, with \textit{the phase space representation} of some operator over $L^2(\mathbb{R}^d)$, which is an operator over a subspace of $L^2(\mathbb{R}^{2d})$. The former characterization is concerned with a specific quantization scheme, while the later with a specific representation of operators in a different quantum state space.

For any operator $A:L^2(\mathbb{R}^d)\rightarrow L^2(\mathbb{R}^d)$, we define its phase space representation as 
\begin{equation}
\mathcal{W}\circ A\circ \mathcal{W}^{-1}:\, \mathfrak{F}^2\rightarrow \mathfrak{F}^2 \ .
\end{equation}

For Heisenberg operators $\hat f={\rm Op}_{H}(f)$, we have
\begin{equation}
\hspace{-20mm}\mathcal{W}\circ \hat f\circ\mathcal{W}^{-1}=\mathcal{W}\circ f\Big(\stackrel{2}{x},-i\hbar \stackrel{1}{\partial _x}\Big)\circ\mathcal{W}^{-1} =f\Big(\stackrel{2}{q}+i\hbar\partial_p,-i\hbar\stackrel{1}{\partial_q}\Big) \ ,
\end{equation}
where we have used the relations $(\ref{eq:rels})$. We use the notation 
\begin{equation}
\check f:=\mathcal{W}\circ \hat f\circ\mathcal{W}^{-1} \ .
\end{equation}

Analogously, the phase space representation of the wave packet quantization of $f$ reads
\begin{equation}
\hspace{-20mm}\mathcal{W}\circ {\rm Op}_{wp}(f)\circ \mathcal{W}^{-1}=\mathcal{P}_{\mathfrak{F}^2}=f\Big(\stackrel{2}{q}+i\hbar\partial_p,-i\hbar\stackrel{1}{\partial_q}\Big)+ O(\hbar^\infty) \ .
\end{equation}

We note that the Heisenberg and wave packet quantizations of the cannonically conjugate pair $q$ and $p$ are identical,
\begin{equation}
\hspace{-25mm}{\rm Op}_{H}(x)={\rm Op}_{wp}(x)=\hat x=m_x \  \ \ {\rm and } \ \ \ {\rm Op}_{H}(p_x)={\rm Op}_{wp}(p_x)=\hat p_x=-i\hbar\partial_x \ ,
\end{equation}
which leads to the phase space representation 
\begin{equation}
\hspace{-25mm}\mathcal{W}\circ{\rm Op}_{H}(x)\circ\mathcal{W}^{-1}=\hat q=q+i\hbar\partial_p \  \ \ {\rm and } \ \ \ \mathcal{W}\circ{\rm Op}_{H}(p_x)\circ\mathcal{W}^{-1}=\hat p=-i\hbar\partial_q \ .
\end{equation}

For a detailed account on the semiclassical wave packet quantization of symplectomorphisms, the thrird essential ingredient of the physical theory next to mixed states and physical quantities, see Nazaikinskii et al \cite{naz1,naz2}.

\subsection{Relation to the Wigner Transform}
\label{wignr}

The wave packet quantization is of no use in the description of systems in strong interaction with their external environment. In this case, a wave fucntion description is not possible, and the quantum system is necessarily described by a statistical ensemble of wave functions. The quantum state of motion is in this case a \textit{density operator}, a trace class rank one projection operator of the form \cite{zachos},
\begin{equation}
\hat \rho =\sum_{\psi\in\mathcal{E}}m(\psi) \mathcal{P}_{\psi} \ ,
\end{equation}
over some countable wave function ensemble, $\mathcal{E}\subset L^2(\mathbb{R}^d)$. $\mathcal{P}_{\psi}$ is the orthogonal projector onto the ray ${\rm span}\{\psi\}$, and $0\leqslant  m(\psi) \leqslant 1$ are probabilities constrained by the normalization $\sum_\psi m(\psi)=1$, reflecting the potential possibilities of the system being in a certain state $\psi$. The notation used $\hat \rho$ does not imply that the density operator is a quantization of some classical phase space density.

In the case of an isolated quantum system, the above reduces to 
\begin{equation}
\hat \rho=\psi\langle \psi,\cdot \rangle_{L^2(\mathbb{R}^d)} \ .
\end{equation}
The phase space representation of the density operator is 
\begin{equation}
\check \rho= \mathcal{W} \circ\hat \rho\circ \mathcal{W} ^{-1}=\Psi\langle \Psi,\cdot \rangle_{\mathfrak{F}^2} \ .
\end{equation}

The Wigner function is
\begin{equation}
\hspace{-20mm}W_\psi(q,p;\hbar)=\bigg(\frac{1}{2\pi\hbar}\bigg)^{d}\int e^{\frac{i}{\hbar}p\cdot x}\psi^\hbar(q-x/2)\bar \psi^\hbar (q+x/2)\,dx \ ,
\end{equation}
the Weyl symbol of the density operator \cite{zachos}. Unlike the wave packet transform, the Wigner transform defines a \textit{bilinear} integral transform of the wave function. It is also real valued, unlike the complex valued phase space Schr\"{o}dinger equation.

The dynamics of the density operator is defined by the Heisenberg equation
\begin{equation}
i\hbar \frac{d\hat \rho }{dt}+[\hat \rho,\hat H]=0\ .
\end{equation}
By taking the Weyl symbol of the left hand side, we have \textit{the von Neumann equation}, or \textit{quantum Liouville equation}. If the underlying classical system is generated by the Hamiltonian $H$, then the time evolution law for the Wigner function reads \cite{zachos}
\begin{equation}
i\hbar \frac{\partial W}{\partial t}=H\star W -W\star H \ .
\end{equation}
Here we have introduced the noncommutative, nonassociative Moyal product \cite{zachos},
\begin{eqnarray}
\hspace{-25mm}f\star g\,(q,p):=\Big(\frac{1}{\pi\hbar}\Big)^{2d}\int\!\!\! \int f(q+\eta,p+\xi) g(q+u,p+v) e^{\frac{2i}{\hbar}(\eta\cdot v-\xi\cdot u)}\, d\eta d\xi du dv \ .
\end{eqnarray}
The Moyal product induces the Moyal bracket, 
\begin{equation}
\{\!\{f,g\}\!\}:=\frac{1}{i\hbar}\Big(f\star g-g\star f\Big) =\{f,g\}+O(\hbar)\ ,
\end{equation}
which is a noncommutative, nonassociative deformation of the Poisson bracket for positive $\hbar$. In terms of the Moyal bracket, the von Neumann equation can be written in a form which makes its characterization as the quantum Liouville equation more apparent, 
\begin{equation}
\frac{\partial W}{\partial t}=\{\!\{H,W\}\!\} \ .
\end{equation}

There is a direct connection between the Wigner formalism and wave packet quantization. The density function $|\Psi|^2$ which defines a phase space probability measure cannot be equated to the Wigner function, as it assumes negative values as well, but rather to the corresponding \textit{Husimi density} \cite{zachos},
\begin{equation}
|\mathcal{W}\psi(q,p)|^2=h_\psi(q,p;\hbar):=\mathcal{G} * W_\psi(q,p;\hbar) \ ,
\end{equation}
the convolution of the Wigner function with a Gaussian, $\mathcal{G}$, over the Heisenberg scale, smearing away oscillations on that scale which sweep both positive and negative signs, rendering it a true phase space densitIy.

The most striking advantage of the Wigner quantization, is that, in contrast to the wave packet quantization, it allows one to consider mixed quantum states. In other words, any phase space wave function in $L^2(\mathbb{R}^{2d})$, along with mild smoothness conditions, are admissible initial data for the von Neumann equation, whereas the phase space Schr\"{o}dinger equation admits only initial data in $\mathfrak{F}^2$, corresponding to pure quantum states, if one is to consider it not merely as a mathematical pseudo-differential equation stripped of all physical content.

\section{Asymptotics of Integrals of Rapidly Oscillating Functions}

In the present we give an outline of the fundamental result on the asymptotic behavior of highly oscillating integrals with complex phase function, of the form resulting from the construction of WKBM wave functions. In particular, a version of the method of stationary phase for complex phase functions.

\subsection{Gaussian Integrals}

For $M\in\mathbb{C}^{n\times n}$ symmetric and ${\rm Re } \,M\succ 0$, and $v\in\mathbb{C}^n$, we have the closed form Gaussian integral
\begin{equation} 
\int e^{-\frac{1}{2}x\cdot M x+v\cdot x}\,dx=\frac{(2\pi)^{n/2}}{\sqrt{\det\, M}}e^{\frac{1}{2}v\cdot M^{-1}v} \ ,
\end{equation}

In the above, as well as throughtout this paper, we take $\sqrt{e^{it}}:=e^{it/2}$ for real $t$, the principal branch of the square root.

\subsection{Almost Analytic Extensions}
\label{almost}

Let $f\in\mathcal{S}(\mathbb{R}^n,\mathbb{C})$. We define its \textit{$N$-analytic extension}, $^{N}\! f:\mathbb{C}^n\rightarrow \mathbb{C}$ as follows \cite{lagrangian}
\begin{equation}
^N\!f(x+iy):=\sum_{k=0}^N\frac{1}{k}\left(iy\cdot \frac{\partial}{\partial x}\right)^kf(x) \ . 
\end{equation}

For arbitrary $N\in\mathbb{Z}_+$, we are led to the concept of \textit{the almost analytic extension}, $\tilde f:= {}^\infty \! f:\mathcal{O}\subset \mathbb{C}^n\rightarrow \mathbb{C}$ such that $\mathbb{R}^n\subset \mathcal{O}$, $\tilde f |_{\mathbb{R}^n}=f$, satisfying the Cauchy-Riemann conditions to all order ($z$ here is not to be confused with the pseudocoordinate introduced in appendix A),
\begin{equation}
\bar \partial_z\tilde f(z)=O\Big(\|{\rm Im}\, z\|^\infty\Big) \ ,
\end{equation}
for $z=x+iy$ nearly real. If $f$ is real analytic, $\tilde f$ coincides with its unique analytic continuation in $\mathcal{O}\times \mathbb{R}^m$.

Following Malther \cite{malther}, a representation of the almost analytic extension of $f$ is 
\begin{equation}
\tilde f(z)=\Big(\frac{1}{2\pi}\Big)^{n/2}\int \varphi(y\cdot \xi)e^{i\xi\cdot z}F(\xi)\,d\xi \ ,
\end{equation}
where $\varphi\in C^\infty_0(\mathbb{R}^n,[0,1])$, supported inside the unit ball $B_1(0)$, becoming unity at a small neighborhood of the origin, $\varphi(0)=1$, and $F$ if the Fourier transform of $f$, (using the same notation as in the appendix for the semiclassical Fourier transform)
\begin{equation}
F(\xi)=\mathcal{F}f(\xi)=\Big(\frac{1}{2\pi}\Big)^{n/2}\int e^{-i\xi\cdot x}f(x)\,dx \ .
\end{equation}

For the sake of convenience we do not use the notation $\tilde f(z,\bar z)$, which formally is more proper, as it stresses the fact that the function $\tilde f$ is not analytic in $z$.

\subsection{Complex Stationary Phase Lemma}
\label{stplemma}

In this subsection we follow the work of Sj\"{o}strand and Melin \cite{sjo}. Firstly, for a smooth function $f:\mathbb{R}^n\rightarrow \mathbb{R}$, we define the stratified manifolds
\begin{equation}
\mathcal{C}_f:=\Big\{x:\, \frac{\partial f}{\partial x}(x)=0\Big\} \ \ {\rm and} \ \ \mathcal{Z}_f:=\Big\{x:\, f(x)=0\Big\} \ ,
\end{equation}
to be \textit{the critical set} and \textit{nodal set} of $f$, respectively.

We now consider the asymptotic behavior of oscillating integrals, which are of the form 
\begin{equation}
I_{a,f}^\hbar(w):=\bigg(\frac{i}{2\pi\hbar}\bigg)^{n/2}\int a(x;\hbar)e^{\frac{i}{\hbar}f(x,w)}\,dx \ ,
\end{equation}
for small positive $\hbar$ and $(x,w)\in\mathbb{R}^n\times \mathbb{R}^m$. The following conditions are assumed:
\\

\noindent 1) the phase function is $f=f_1+if_2\in C^\infty (U_n\times V_m,\mathbb{C})$, where $U_n$ and $V_m$ are neighborhoods of the origins of $\mathbb{R}^n$ and $\mathbb{R}^m$ respectively, with nonnegative imaginary part, $f_2\geqslant 0$, equality being saturated at the origin $(x,w)=0$; the origin is also a nondegenerate stationary point, $\partial _xf(0)=0$ with $\det\,f''_{xx}(0)\neq 0$, which leads to ${\rm Im}\, f''_{xx}(x,w)\succ 0$ in $(U_n\times V_m)$.
The critical manifold of $f_2$, $\mathcal{C}_{f_2}:=\Big\{(x,w):\, \partial_xf_2(x,w)=0\Big\}$, is thus nondegenerate. Additionally, the nodal set $\mathcal{Z}_{f_2}:=\Big\{(x,w):\, f_2(x,w)=0\Big\}$ is identified with the critical set $\mathcal{Z}_{f_2}\equiv \mathcal{C}_{f_2}$.
\\

\noindent 2) The amplitude is $a\in S^0_{1-\delta}(\mathbb{R}^n\times \mathbb{R}_+,\mathbb{C})$, where $\delta<\frac{1}{2}$, supported in a neighborhood of the origin within $U_n\times V_m$, admitting a regular semiclassical asymptotic expansion in the interior of ${\rm supp}\, a(\, \cdot \,;\hbar)$,  
\begin{equation}
a(x;\hbar)\sim \sum_{k=0}^\infty\hbar^ka_k(x) \ , \ \ \hbar\rightarrow 0^+ \ .
\end{equation}

As $f$ is smooth, it admits an almost analytic extension $\tilde f$ into the complex domain $\mathcal{O}\times\mathbb{R}^m$, where $\mathcal{O}\subset \mathbb{C}^n$ containing the real subspace $\mathbb{R}^n$, satisfying the Cauchy-Riemann conditions to all orders,
\begin{equation}
\bar \partial_z\tilde f(z,w)=O\Big(\|{\rm Im}\, z\|^\infty\Big) \ ,
\end{equation}
for $z=x+iy$ nearly real.

Let there be a real critical point, i.e., $z=x_0+i0$ such that $\partial_z\tilde f(x_0,w)=0$. By the implicit function theorem, the critical manifold $\mathcal{C}_{\tilde f}$ continues locally into the complex domain $\mathcal{O}\times \mathbb{R}^m$, parameterized by the almost analytic coordinate chart 
\begin{equation}
z=\underline z(w)\ , 
\end{equation}
always in a small neighborhood of $(x_0,w_0)$, with $x_0=\underline z(w_0)$, say the ball $B_{\delta}(x_0,w_0)$ for some $0<\delta<1$. The above is termed an \textit{almost solution} of $\partial_x f(x,w)=0$.

The positivity of the imaginary part continued into the complex critical manifold is guaranteed by Sj\"{o}strand \cite{sjo},
\begin{equation}
{\rm Im}\, \tilde f(z,w)\geqslant c\,({\rm Im}\, z)^2 \ ,
\end{equation}
for $(z,w)\in \mathcal{C}_{\tilde f}\cap B_{\delta}(x_0,w_0)$, and some $c>0$. By notation, it should be clear that ${\rm Im}\,\tilde f$ does not stand for $\tilde f_2$. Additionally, by the G\"{a}rding inequality, we have 
\begin{equation}
\|\nabla_{(x,y)} {\rm Im}\, \tilde f(z,w)\|^2_{\mathbb{C}^n}\leqslant c\,{\rm Im}\, \tilde f(z,w) \ ,
\end{equation}
for some different $c>0$.

By Sj\"{o}strand (\cite{sjo}, p. 145), $\nabla_{(x,y)}{\rm Im}\,\tilde f=0$ is equivalent to $\partial_z\tilde f=0$. Thus, the later inequailty implies that points belonging to the nodal set of the imaginary part of $\tilde f$, $(z,w)\in \mathcal{Z}_{{\rm Im}\,\tilde f}$, are necessarily critical points; thus, we have
\begin{equation}
\mathcal{Z}_{{\rm Im}\, \tilde f}\subset \mathcal{C}_{\tilde f} \ .
\end{equation}
However, by the bound for the imaginary part of $\tilde f$, we have that any point $(z,w)\in \mathcal{C}_{\tilde f}\cap B_{\delta}(x_0,w_0)$ with ${\rm Im}\, z\neq 0$, cannot belong to the nodal set $\mathcal{Z}_{{\rm Im}\, \tilde f}$, and thus, by the connectedness of $\mathcal{Z}_{{\rm Im}\,\tilde f}$, all zeroes of ${\rm Im}\, \tilde f$ are necessarily real critical points,
\begin{equation}
\mathcal{Z}_{{\rm Im }\,\tilde f}\equiv \mathcal{Z}_{{\rm Im }\, f}\subset \mathbb{R}^n\times \mathbb{R}^m \ .
\end{equation}

One must show that all real stationary points are necessarily zeroes of ${\rm Im}\, \tilde f$, i.e., that the real restriction $\mathcal{C}_{\tilde f}\cap \mathbb{R}^n\times \mathbb{R}^m$ is identified with $\mathcal{Z}_{{\rm Im}\, f}$ as manifolds in $\mathbb{R}^n\times \mathbb{R}^m$. This requirement is met by the class of WKBM states as defined in the previous sections.
\\

\noindent \textbf{Theorem} (Sj\"{o}strand and Melin \cite{sjo}) \textit{Let $a$ and $f$ be amplitude and phase functions as defined above. Then, there is a neighborhood $U_n$ and $V_m$ of the origins of $\mathbb{R}^n$ and $\mathbb{R}^m$ respectively, and differential operators $L_k(w,\partial_z)$ of order less than $2k$, with smooth coefficients in $w\in V_m$, such that}
\begin{eqnarray}
\hspace{-20mm}\bigg(\frac{i}{2\pi\hbar}\bigg)^{n/2}\int a (x;\hbar)e^{\frac{i}{\hbar}f(x,w)}\,dx-\frac{\tilde a_0(\underline z(w))}{\sqrt{\det(-\tilde f''_{zz}(\underline z(w),w))}}e^{\frac{i}{\hbar}\tilde f(\underline z(w),w)}\nonumber\\
\sim\sum_{k=1}^\infty\hbar^k \Big(L_k(w,\partial_z)\tilde a_k\Big)(\underline z(w)) \ , \ \ \hbar\rightarrow 0^+ \ ,
\end{eqnarray}
\textit{in $S^{-n/2}_{0,1}(V_m\times \,]0,1])$. In the above, $z=\underline z(w)$ is the equation defining the almost analytic manifold $\mathcal{C}_{\tilde f}$ in a complex neighborhood of $U_n\times V_m$, $\tilde f$ and $\tilde a_0$ are the almost analytic extensions of $f$ and $a_0$ in the same complex neighborhood. The square root in the left hand side is the principal branch.}

In the above context, we call a point $z_0$ a \textit{real stationary point} if there is a $w\in\mathbb{R}^m$ for which the following equations hold
\begin{eqnarray} 
\partial_x f(z_0,w)=0 \nonumber \\
{\rm Im}\, f(z_0,w)=0 \ ,
\end{eqnarray}
which, of course, implies that $z_0$ is itself real. We are lead to define \textit{the real critical manifold} of $f$ as 
\begin{equation}
\mathcal{C}^{\mathbb{R}}_f:=\Big\{(x,w):\, \partial_xf(x,w)=0 \ , \ \ {\rm Im}\,f(x,w)=0\Big\} \ .
\end{equation}

It is clear that for oscillating integrals of the above form, real stationary points -if any- give a non exponentially decaying contribution in $\hbar$, while genuinely complex ones will give exponentially small contributions in $\hbar$.

The solution of $\partial_x f(x,w)=0$, which determines the stationary manifold, is $x=\underline z(w)$. For $w$ close enough to the real stationary manifold, $\mathcal{C}_f^\mathbb{R}$, we have $\underline z(w)=\underline x(w)+\varepsilon(w)$, where $\underline x(w)$ parameterizes the real manifold at the point projected down by ${\rm Re}\, z(w)$, and $\varepsilon_w$ (which is not purely imaginary) is linear in the distance $d_w={\rm dist}\,\Big(w,\mathcal{C}_f^\mathbb{R}\Big)$, for $w$ close enough to the manifold. 

As, by definition, $\underline x(w)\in\mathcal{C}_f^\mathbb{R}$, near the real manifold, the phase reads
\begin{equation}
f(\underline z(w),w)=f_1(\underline x(w),w)+\frac{1}{2}\varepsilon_w\cdot f''_{xx}(\underline x(w),w) \varepsilon_w+O(d_w^3) \ .
\end{equation}

The real critical manifold approximation amounts to 
\begin{equation}
I_{a,f}(w;\hbar)\sim \frac{a_0(\underline x(w))}{\sqrt{\det(-f''_{xx}(\underline x(w),w))}}e^{\frac{i}{\hbar} f(\underline x(w),w)} \ ,
\end{equation}
where $\underline x(w)$ is a parametrization of the real critical manifold $\mathcal{C}_f^\mathbb{R}$. The above approximation gives the same leading behavior for $I_{a,f}^\hbar(w)$ as the above almost analytic extension expansion. There is no need to employ the almost analytic extension of $f$, as $\tilde f(x,w)=f(x,w)$ for real $x$. 
The branch of the root $\sqrt{\det-f''_{xx}}$ is chosen so that $Arg(\det(-f''_{xx}))= \sum_{k=0}^n Arg{\lambda_k}$, where  ${\lambda_k}$ are the eigenvalues of  the Hessian $\det(-f''_{xx})$,  with $Arg{\lambda_k} \in (-3\pi/2, \pi/2)$.

The quadratic term contributes $O\bigg(\frac{1}{\hbar}\varepsilon_w\cdot f''_{xx}(\underline x(w),w) \varepsilon_w\bigg)=O(1)$, as the microscopic vicinity of $\mathcal{C}_f^\mathbb{R}$ is such that $\|\varepsilon_w\|_{\mathbb{C}^n}\asymp d_w\asymp \sqrt{\hbar}$. It is clear that in the case there are no real critical points $\underline x(w)=0$, the imaginary part of the phase is nonzero, and thus there is an overall exponentially decaying behavior. Explicitly, if $\mathcal{C}_f^\mathbb{R}=\emptyset$,
\begin{equation}
I_{a,f}(w;\hbar)=O_w\Big(e^{-\frac{1}{\hbar}{\rm Im}\,\tilde f(\underline z(w),w)}\Big) \ , \ \ \hbar\rightarrow 0^+ \ ,
\end{equation}
the integral is exponentially decaying everywhere in $\mathbb{R}^m$. \\

\newpage

{\bf Bibliography}
\\

\end{document}